\documentclass[journal,comsoc,10pt]{IEEEtran}
\usepackage[export]{adjustbox}
\usepackage{textcomp}
\usepackage{algorithm}
\usepackage{algpseudocode}
\usepackage{tabularx}
\usepackage[table]{xcolor}
\usepackage{booktabs}
\usepackage{array}
\newcolumntype{P}[1]{>{\centering\arraybackslash}p{#1}}
\newcolumntype{M}[1]{>{\centering\arraybackslash}m{#1}}
\usepackage{multicol}
\usepackage{multirow}
\usepackage{caption}
\usepackage{subcaption}
\usepackage{color}
\usepackage{capt-of}
\usepackage{balance}
\usepackage{dsfont}
\usepackage{graphicx}
\usepackage{epstopdf}
\usepackage{stfloats}
\usepackage[cmintegrals]{newtxmath}
\usepackage{amsfonts}
\usepackage{times}
\usepackage{latexsym}

\usepackage{amssymb}
\usepackage{amsmath}
\usepackage{verbatim}
\usepackage{bm}
\usepackage{cancel}
\usepackage{wrapfig}
\usepackage{listings} 
\usepackage{url}
\usepackage{xcolor}
\usepackage{lettrine}
\usepackage{cite}
\usepackage{pifont}
\usepackage{diagbox}
\newtheorem{theorem}{Theorem}

\newtheorem{remark}[theorem]{Remark}

\renewcommand{\IEEEQED}{\IEEEQEDclosed}
\graphicspath{{figure/}{figures/}}
\IEEEoverridecommandlockouts

\begin{document}
\title{Reinforcement Learning-based Joint User Scheduling and Link Configuration in Millimeter-wave Networks}
\author{Yi Zhang, \IEEEmembership{Member,~IEEE} and Robert W. Heath Jr., \IEEEmembership{Fellow,~IEEE}
\thanks{Manuscript received Sept 2021; revised April 2022; revised Aug. 2022, accepted Oct. 2022. This work was partially supported by the U.S. Army Research Office under grants W911NF1910221, NSF-CNS-1731658, and NSF-CCF-2225555. 
The associate editor coordinating the review of this article and approving it for publication was Dr. Vincenzo Sciancalepore. 
Yi Zhang is with The University of Texas at Austin, TX 78731, US (e-mail: yi.zhang.cn@utexas.edu). Robert W. Heath Jr. is with North Carolina State University, Raleigh, NC 27695, US (e-mail: rwheathjr@ncsu.edu).}}
\markboth{}
{}
\maketitle

\begin{abstract}
In this paper, we develop algorithms for joint user scheduling and three types of mmWave link configuration: relay selection, codebook optimization, and beam tracking in millimeter wave (mmWave) networks. Our goal is to design an online controller that dynamically schedules users and configures their links to minimize system delay.
To solve this complex scheduling problem, we model it as a dynamic decision-making process and develop two reinforcement learning-based solutions.
The first solution is based on deep reinforcement learning (DRL), which leverages the proximal policy optimization to train a neural network-based solution. 
Due to the potential high sample complexity of DRL, we also propose an empirical multi-armed bandit (MAB)-based solution, which decomposes the decision-making process into a sequential of sub-actions and exploits classic maxweight scheduling and Thompson sampling to decide those sub-actions.
Our evaluation of the proposed solutions confirms their effectiveness in providing acceptable system delay. It also shows that the DRL-based solution has better delay performance while the MAB-based solution has a faster training process.
\end{abstract}


\section{Introduction}\label{sec:Intro}
\subsection{Motivations}\label{subsec:Motivation}
MmWave is a key feature of next-generation communication systems due to the high bandwidths available versus lower frequency~\cite{Ted_mmWave_will_work_2013Access}. 
Large-scale phased arrays with compact size, coupled with the hybrid architecture, provide higher antenna gain and flexible MIMO operations for mmWave link configuration~\cite{Han_LargeScaleMmWaveComMag14}.
Configuring these links is challenging, especially when considering user scheduling and the rapidly changing channels due to mobility~\cite{Multinomial_TS_BeamTracking_Aykin_Infocom20}.
We summarize three of the most important challenges below:

\subsubsection{Joint user scheduling and relay selection}
Dynamic and static obstacles may block direct line-of-sight (LOS) paths, which provide high-quality mmWave links~\cite{Relay_selection_scheduling_MASS2017}. These blockages may create a long outage duration~\cite{Outage_duration_TVT2021}.
This limits the network coverage and capacity.
One practical solution is to leverage relay-assisted transmission, by enabling device-to-device (D2D) links, to bypass the blockages and extend the network coverage~\cite{Relay_mmWave_ComMag2017}.
Joint user scheduling and relay selection are challenging, though, in fast-varying mmWave networks. 
First of all, it is unrealistic to probe all relays (users in the network) to decide on an instantaneously optimal relay due to large beam training overhead~\cite{Relay_probing_MmWave_Wei_TVT16}.
Second, potential imbalanced data traffic demands among users play a role in scheduling users and relays~\cite{Blockage_scheduling_WPAN_Niu_TVT15}.
For example, without a significant quality of service (QoS) degradation, 
an access point (AP) could frequently select the user with low data traffic demand as a relaying node for the data transmission to the users that are far away or suffering from blockage loss~\cite{Blockage_scheduling_WPAN_Niu_TVT15}. 
Excessively selecting a single user (especial the user with the high demand) as a relay, however, greatly impacts its QoS.
Thus, it is imperative to design a low-overhead algorithm that jointly performs user scheduling and relay selection without sacrificing the QoS of users.
 
\subsubsection{Codebook optimization with beamforming gain, training overhead, and link robustness}
Codebook-based exhausted/hierarchical beam scanning was adopted by IEEE 802.11ad/ay~\cite{802.11ad_standard:2016,11ay_standard_2019} 
and 5G NR~\cite{5G_NR_standard_V16:2019}.
The large antenna arrays are used to generate highly directional beams, hence providing beamforming gain in LOS channels.
The gain, however, comes with beam scanning overhead that grows when a large number of narrow beams are used~\cite{MUTE_Yasaman_Mobicom18}
(even with hierarchical search at the very beginning stage).
In addition to the overhead, a mmWave link that is composed of narrower beams
is more vulnerable to beam misalignment (or even the outage) 
caused by device mobility, as depicted in Fig.~\ref{fig:outage_model}.
It has been shown that even a relatively small beam alignment error
could lead to significant rate reduction~\cite{ThoHea:Ergodic-Rate-AdHoc:18,Lee_BA_UE_scheduling_with_Mobility_Wiopt19}.
As a result, selecting an appropriate codebook that resolves this tension among beamforming gain, training overhead, and link robustness is critical. 
In particular, a dynamic approach is encouraging since the optimal codebook is generally user-dependent and scenario-dependent~\cite{TSCB_YiZhang_Mobihoc_2021,V2X_mmWave_measurements_Mobicom20}.

\subsubsection{Impact of beam training periodicity on user scheduling and beam training efficiency}
5G NR specifies five possible training periods that vary from 10 ms to 160 ms~\cite{5G_NR_standard_V16:2019}. The selection of the actual value is left to the vendors.
On the one hand, a smaller training period would be favorable and flexible for maintaining the links with users that are fast-moving or located close to the transmitter~\cite{V2X_mmWave_measurements_Mobicom20},
which may however reduce the beam training efficiency.
On the other hand, a longer training period is more efficient as it has a longer data transmission duration after each beam training, but it may waste time resources when the connected user has low data traffic demands, hence adding unnecessary delay to the other users.
One potential solution is to use a smaller training period attached with an optional beam tracking period whose duration is the same as the training period.
In other words, it is analogous to dividing a long training period (e.g. 160 ms) into multiple flexible slots (e.g. 16 slots of 10 ms). 
Some slots could be configured as beam tracking periods for users with high traffic demands. 
This approach avoids allocating a long period to a single user all at once but with the cost of beam tracking overhead, which is relatively small and even negligible. 
Though, it is still unclear how frequent or when beam tracking should be applied.

\subsection{Contributions}\label{subsec:contribution} 
In this paper, we design a controller that jointly schedules users and configures mmWave links in a multi-user mmWave network under mobility. We consider three types of link configuration: relay selection, codebook selection, and beam tracking. Our goal is to minimize the average packet delay. In particular, it turns out that maximizing the system throughput can greatly reduce the delay.
Due to the large overhead associated with channel probing, it is unrealistic to provide a deterministic optimal strategy~\cite{Relay_probing_MmWave_Wei_TVT16}. In contrast, an online solution is a good fit for the stochastic nature of mmWave networks under mobility. 
Recently, reinforcement learning (RL) has been shown to be a powerful tool to solve complex decision-making problems using dynamic programming~\cite{RL_Intro_Sutton_book18,DRL_wireless_overview_VTMag19,Luong_DRL_wireless_ComSurvey_19} (please see Sec.~\ref{sec:subsec:related_work} for more examples of applying RL in wireless system designs).
In this work, we exploit RL to jointly solve user scheduling and mmWave link configuration problem.
The contributions of our paper are summarized below:

\subsubsection{System Modeling} 
We model the joint user scheduling and link configuration of a mmWave network as a queuing system, where each queue buffers the data for a user. Three types of mmWave link configuration are considered, i.e. relay selection, codebook selection, and beam tracking. We mathematically formulate the system as a discrete-time decision-making process by showing how the queues are influenced by the decisions on link configuration, under various environmental dynamics, such as network traffic, device mobility, blockage, and channel fading. Prior work has not jointly considered these configurations with user scheduling.

\subsubsection{DRL-based controller} 
We transform the modeled system into a partially observable Markov decision process (POMDP) and further propose a \textit{DRL-based controller} to dynamically schedule the users and configure their mmWave links.
In particular, we exploit the proximal policy optimization (PPO)~\cite{PPO_Schulman_17} within the advantage-to-critic (A2C) framework~\cite{A3C_Mnih_ICML16} to train a neural network (NN)-based stochastic controller,
which is carefully crafted such that the interplays among user scheduling and link configuration are included.

\subsubsection{Empirical MAB-based controller} 
To combat the potential large sample complexity brought by the DRL-based solution, 
we propose a sample-efficient and low-complexity learning algorithm, called \textit{Empirical MAB-based controller}.
This algorithm decomposes the entire decision-making process into a sequence of four sub-actions: namely that user scheduling, selecting relay, choosing codebook, and deciding on beam tracking. It then exploits maxweight scheduling~\cite{MaxWeight_TIT_93} and bandit algorithms~\cite{TSCB_YiZhang_Mobihoc_2021} to sequentially decide these four sub-actions.

\subsubsection{Comprehensive empirical evaluation} 
We evaluate the two proposed RL-based controllers by numerical simulation with practical system parameters. Our results show that both of them are effective in providing acceptable system delay. In particular, the DRL-based solution provides better system performance while the MAB-based solution results in a faster training process.

\subsection{Related work}\label{sec:subsec:related_work}
\subsubsection{RL-based user scheduling}
In queue-based scheduling, the classic maxweight algorithm~\cite{MaxWeight_TIT_93} 
is throughput optimal and provides good delay performance~\cite{Delay_analysis_maxweight}, though requires the channel state information beforehand.
The channel in practical mobile networks is dynamic, which drives the need for learning scheduling policies~\cite{DDPG_Sch_Min_Delay_LTE_PIMRC_2020}. Some recent work has been exploiting the newfound power of RL techniques~\cite{RL_Intro_Sutton_book18}, e.g deep Q-network (DQN)~\cite{DQN_Mnih_Nature15} and deep deterministic policy gradient (DDPG)~\cite{DDPG_Lillicrap_ICLR16}.
In~\cite{LSTM_DQN_Min_Delay_Globecom_2018,DQN_RNN_Min_Delay_Globecom_2020}, two DQN-based DRL schedulers were developed
to minimize the system latency. 
In~\cite{Intel_DRL_OFDMA_RA_Globecom19, DDPG_Sch_Min_Delay_LTE_PIMRC_2020}, two DDPG-based DRL schedulers were further developed 
to allocate the time/frequency resource allocation among users to minimize the queuing delay.
Besides, the DDPG algorithm was exploited to perform wireless routing in~\cite{Gupta_DRL_link_schedule_TWC20} to learn a scheduling policy that minimizes the end-to-end packet delay. 
This prior work shows that their obtained delay performance is close to or even superior to the maxweight scheduling. 
Unfortunately, either user mobility or the effect of mmWave link configuration (e.g. relay selection and beam tracking) were considered in the user scheduling in~\cite{DDPG_Sch_Min_Delay_LTE_PIMRC_2020,LSTM_DQN_Min_Delay_Globecom_2018,DQN_RNN_Min_Delay_Globecom_2020,Intel_DRL_OFDMA_RA_Globecom19,Gupta_DRL_link_schedule_TWC20}. 
In contrast, we study user scheduling by considering multiple types of mmWave link configuration and exploit the most recently proposed PPO algorithm for the learning process rather than using DQN or DDPG.

\subsubsection{Relay selection in mmWave systems}
MmWave relaying schemes have been intensively studied (see~\cite{Relay_probing_overview_mmWave_Access2020} and references therein).
For example, relay selection and spatial reuse were jointly optimized to reduce the system delay of a mmWave WPAN in~\cite{Blockage_scheduling_WPAN_Niu_TVT15}, but only for static channel conditions. 
A relay probing strategy was investigated in~\cite{Relay_probing_MmWave_Wei_TVT16}, 
but the impact of the relay selection on the user scheduling was not considered. 
Motivated by the fact that relay selection is resource-demanding under rapidly-varying mmWave channel~\cite{Predictive_Prelay_selection_ACCESS2019},
some recent work has started leveraging RL to perform dynamic selection rather than solving instantaneous optimization problems as~\cite{Blockage_scheduling_WPAN_Niu_TVT15,Relay_probing_MmWave_Wei_TVT16} did.
For example, in~\cite{Zhang_DRL_relay_power_selection_WCL20}, a DQN-based relay selection was developed for a downlink vehicular-to-infrastructure network. In~\cite{SCAROS_epsilon_greedy_backhauling_routing_JSAC_2019}, the $\epsilon$-greedy policy was used to 
solve the routing selection in a self-backhauled mmWave cellular network.
Both~\cite{Zhang_DRL_relay_power_selection_WCL20} and~\cite{SCAROS_epsilon_greedy_backhauling_routing_JSAC_2019} showed that their RL-based solutions do not require prior information on network dynamics but their delay performance is robust to complex environmental dynamics.
Nevertheless, the applications of RL in the mmWave band are still largely open.
In contrast to prior work, we study the mmWave relay selection under a complicated scenario by considering various factors, such as codebook selection, beam tracking design, and data requirements of users.

\subsubsection{MmWave codebook optimization} 
The codebook optimization problem was initially investigated in~\cite{MU_BW_opt_Hossein_ICC_2015, Liu_JSAC_Wiopt_2019} by optimizing beamwidth.
Their solutions, however, depend on prior knowledge such as channel state information, which restricts their practical use in mobile networks where the channels are usually rapidly changing. In contrast, our proposed RL-based solutions are model-free, and thus more flexible for deployments at different sites.
Recently, data-driven approaches have been widely exploited.
A set of beam pairs was learned via deep learning in~\cite{DL_Learning_beam_pairs_Wang_Access_2019} and a geo-located context database was built in~\cite{FilSciDevCap_GeoDataBase_TMC} to assist the beam width/direction selection. These two offline approaches, however, require a large amount of data for a given site. This would limit their fast implementation when the system dynamics are high.
Furthermore,~\cite{Site_codebook_learning_Wang_Heath_TWC21},~\cite{Jeong_OnlineTracking_CBSelection_Infocom20}, and~\cite{alrabeiah_NN_beam_codebook_2020} designed online learning-based algorithms to adaptively optimize codebook beam patterns.
In~\cite{Gao_DRL_bw_powerOptimization_CL20}, a joint beamwidth and power control problem was solved via DRL.

\subsubsection{Beam tracking design}
Beam tracking has been widely studied for a single user. In~\cite{Palacios_Tracking_mmWaveChannel_Infocom17}, hybrid mmWave phased arrays were used to perform beam tracking by probing the channel in multiple spatial directions simultaneously.
In~\cite{WidebandChannelTracking_Nuria_TWC2021}, the sparsity of the mmWave channel was exploited to perform wideband channel tracking. 
The effects of beam tracking and data communication durations were optimized in~\cite{Shahram_Robust_BeamTracking_WiOPT19}.
A TS-based beam tracking algorithm
was proposed in~\cite{Multinomial_TS_BeamTracking_Aykin_Infocom20},
where the time-varying optimal beam is tracked with numerous feedbacks before each beam retraining.
The prior work, however, focused on beam tracking design for the single-user case, whereas the potential delay introduced by tracking to other users was not considered. In contrast, our work focuses on learning \textit{whether and when} the beam tracking should be performed in a multi-user mmWave network such that the system delay is minimized.

\begin{figure*}[t!]
	\begin{subfigure}[t]{0.45\textwidth}
		\centering
		\includegraphics[width = 0.75\textwidth]{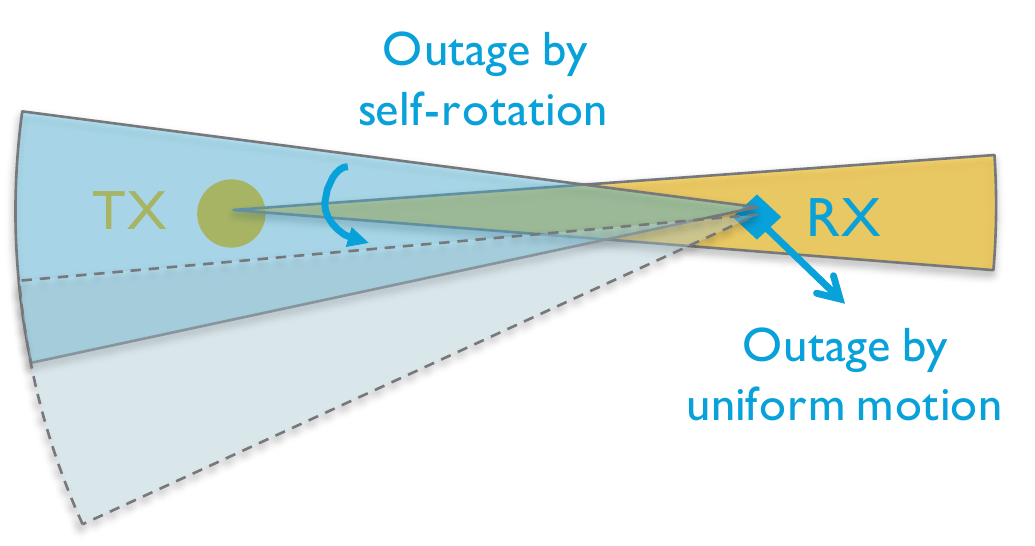}
		\captionsetup{type=figure,justification=centering}
		\caption{}
		\label{fig:outage_model}
	\end{subfigure}
	\begin{subfigure}[t]{0.55\textwidth}
		\centering
		\includegraphics[width = 0.99\textwidth]{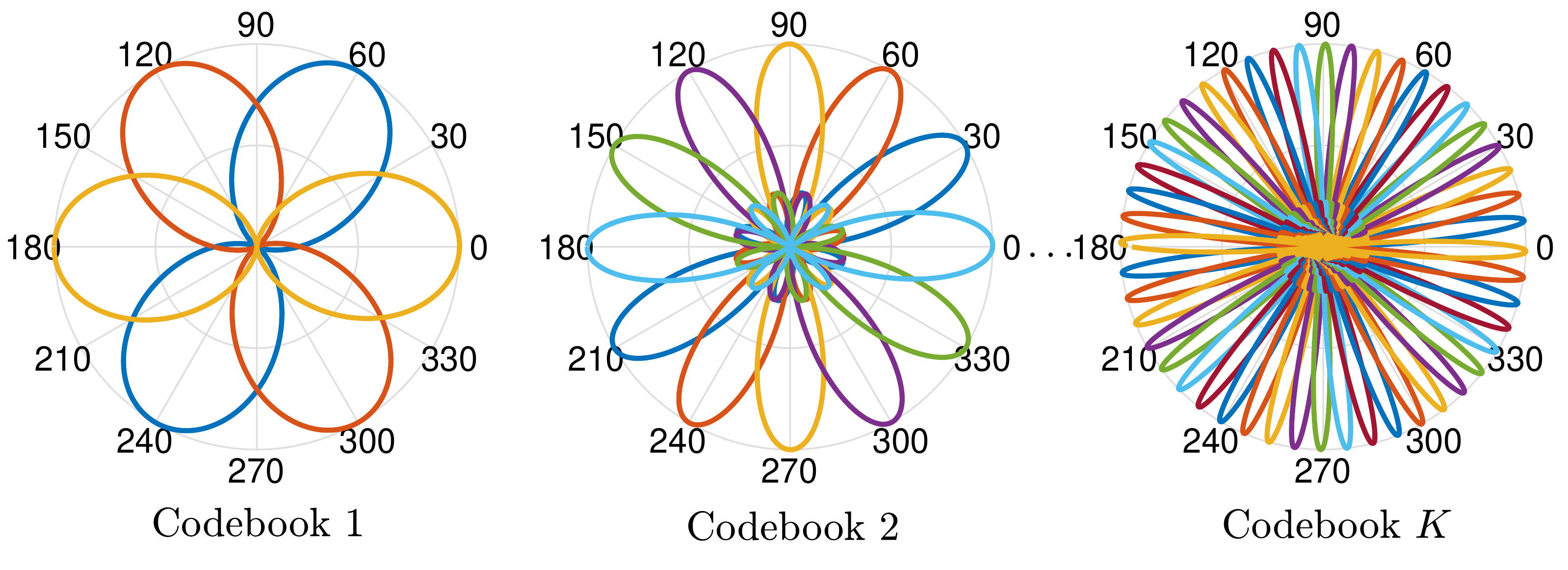}
		\captionsetup{type=figure,justification=centering}
		\caption{}
		\label{fig:codebook}
	\end{subfigure}

	\caption{Codebooks of different beam resolutions: (a) When using narrower beams, a link outage is more likely to incur within a time slot due to self-rotation or uniform motion of device. (b) Pictorial examples of $K$ codebooks of different beamwidth.}
\end{figure*}

\begin{figure*}[t!]
	\begin{subfigure}[t]{0.64\textwidth}
		\centering
		\includegraphics[width = 0.99\textwidth]{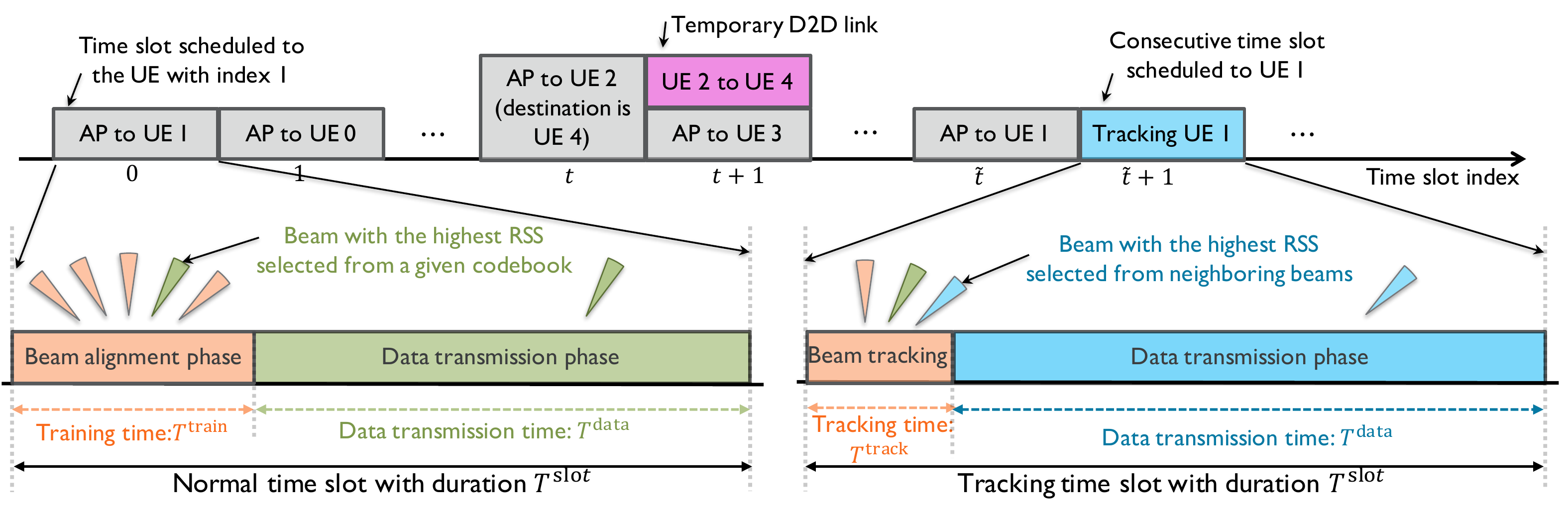}
		\captionsetup{type=figure,justification=centering}
		\caption{Decomposition of a communication time slot}
		\label{fig:time_slot}
	\end{subfigure}
	\begin{subfigure}[t]{0.34\textwidth}
		\centering
		\includegraphics[width = 0.99\textwidth]{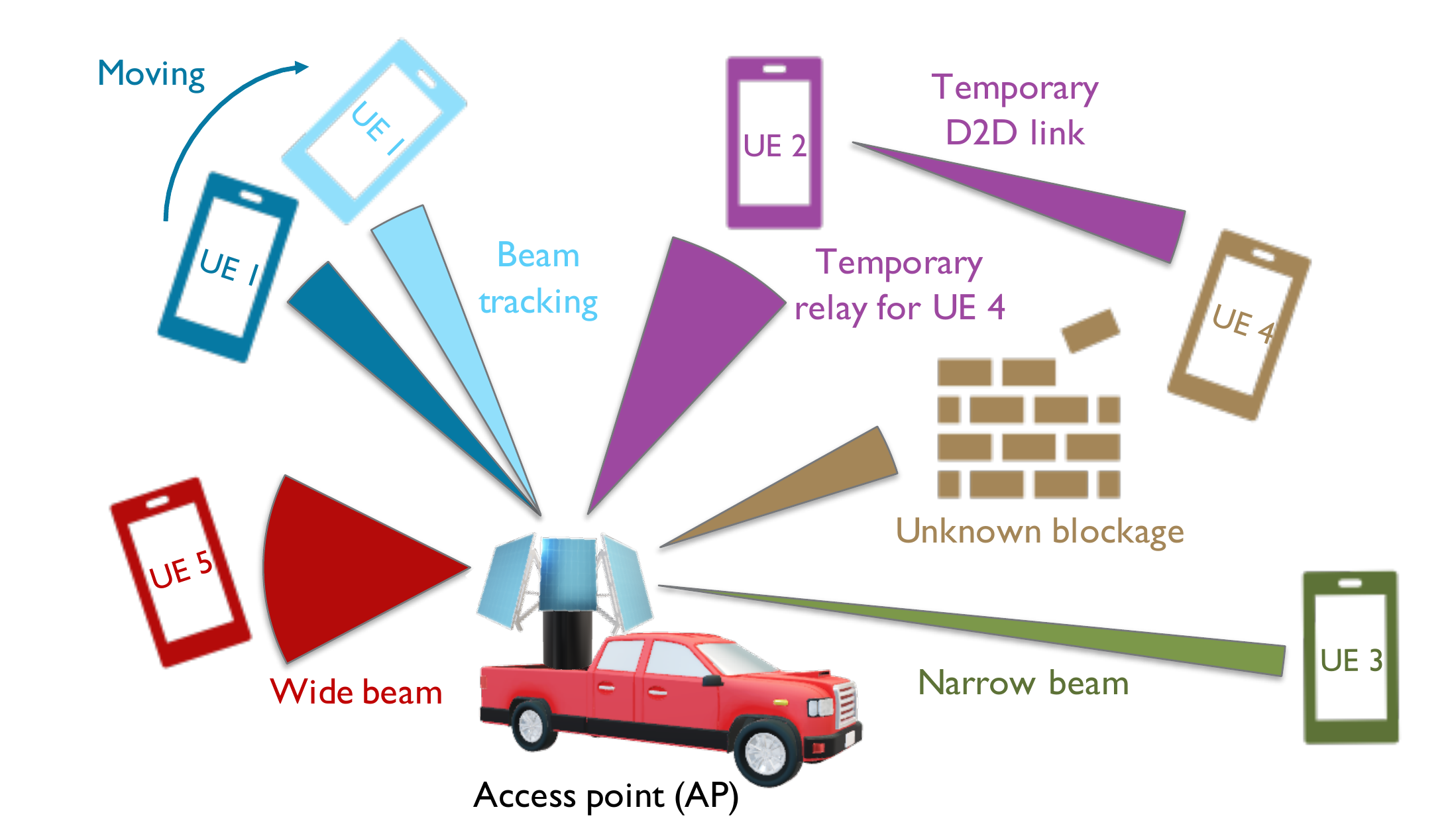}
		\captionsetup{type=figure,justification=centering}
		\caption{MmWave link configuration}
		\label{fig:network_model}
	\end{subfigure}
	\caption{User scheduling and link configuration in mmWave network}
	\label{fig:network_model_and_time_slot}
\end{figure*}

\subsection{Notation}
We use the bold lowercase font to represent vectors and the normal font
for scalars. 
We use $q_i$ to denote the $i$-th element of a vector $\mathbf{q}$.
We denote $[K]^+$ as the set $\left\{1,\hdots,K\right\}$, and $[K]$ as the set $\left\{0\right\}\cup[K]^+$.
$\mathds{1}\left\{\cdot\right\}$ denotes the indicator function.
$\mathbf{1}$ is an all-ones vector with a context-dependent dimension. 
$\lceil\cdot\rceil$ and $\lfloor\cdot\rfloor$ denote the ceiling and flooring operator, respectively. 
$\text{Mod}\left(\cdot\right)$ denotes the Modulo operation.
$X\sim \text{Bernoulli}\left(p\right)$ means that $X$ follows a Bernoulli distribution with parameter $p$.
$Dir\left(\alpha_{0},\hdots,\alpha_{M}\right)$ 
denotes a Dirichlet distribution with parameter vector $ [\alpha_{0},\hdots,\alpha_{M}]^{\text{T}}$.

\section{System model}\label{sec:sys_model}
We consider downlink transmission in a mmWave network, in which an AP (or base station) communicates with $U$ \textit{mobile} users (UE) in a slot-based manner. 
For each time slot (of a fixed duration $T^\text{slot}$), the AP dedicates all bandwidth $B$ to a single UE. The time slot is composed of a beam alignment (BA) phase and a data transmission (DT) phase, as shown in Fig.~\ref{fig:time_slot}. During the BA phase, the AP performs a 2D codebook-based beam training, where a codebook refers to a collection of directional beams that share the same beamwidth and together cover the whole spatial space, as shown in Fig.~\ref{fig:codebook}.
The AP and UE test all the beams in their receptive codebooks in the BA phase and the indexes of the best transmitting-receiving beam pair along with the received signal strength (RSS) (or estimated SNR) is feedback to the AP.
Afterward, the AP and UE would use this identified best beam pair to send/receive data,
which is referred to as the DT phase. In particular, the highest supportable modulation and coding scheme (MCS) would be used based on a predefined RSS-MCS (or SNR-MCS) table. We refer to the $u$-th UE in the network as UE $u$ or device $u$, where the integer $u$ is its network ID. Without any ambiguity, we assign ID 0 to the AP for notational simplicity.
Three types of link configuration are available in the system. They are codebook selection, relay selection, and beam tracking. A pictorial example is given in Fig.~\ref{fig:network_model}, which is described below:

\subsubsection{Codebook selection}
To be flexible with different channel conditions and locations of UEs, 
$K$ codebooks of different beamwidth are available at the AP while only a single codebook (the one with the widest beamwidth) is used by UEs since UEs are generally equipped with smaller arrays, implying fewer antennas and wider beams. Hence, codebook selection is only enabled for the AP.
In this work, we focus on the codebooks shown in Fig.~\ref{fig:codebook}.
The generation of codebooks with different beamwidth is out of the scope of this work and a potential approach is to use the antenna on/off techniques~\cite{CB_Design_Xiao_TWC_2017,SANBA_YiZhang_Mobihoc_2019}.
It is worth pointing out that any beam training strategies or a strategy with different parameters (including hierarchical search) can be incorporated into our codebook selection framework by regarding them as different ``abstract codebooks"~\cite{TSCB_YiZhang_Mobihoc_2021}. 

\subsubsection{Relay selection}
An example of relay-assisted transmission is given in Fig~\ref{fig:network_model_and_time_slot}: the LOS path between UE 4 and the AP is temporarily blocked. The AP first transmitted data (with UE 4 as destination) to UE 2 and 
in the consecutive time slot, UE 2 forwards its received data to UE 4.
We refer to the link between AP and any UE as a \textit{main link} 
while the link between two UEs as a \textit{D2D link}. 
As a result, a UE could be in one of the following four modes: 
idle, being a receiver (RX) in a main link, being a transmitter (TX) in a D2D link, 
or being a RX in the D2D link. 
We consider that all devices (AP and UEs) are working in a half-duplex mode, 
which implies that any devices cannot simultaneously participate in 
a main link and a D2D link.

\subsubsection{Beam tracking} 
When the AP continues serving the same UE in consecutive time slots, testing the neighboring beams of the current best beam pair would be sufficient to maintain the link quality.
This process is called beam tracking. A consecutive time slot that tracks a UE is called a \textit{tracking slot}.
In our studied system, beam tracking is not activated to track a relay.

In the following subsections, we will characterize the studied system by mathematically modeling network traffic, mobility, blockage, and mmWave link configuration. 
The notation is summarized in Table~\ref{tab:notations} for reference.
\begin{table}[t!]
    \renewcommand*{\arraystretch}{1.2}
    \captionsetup{type=table,justification=centering}
    \captionof{table}[t]{List of Key Notations} 
    \label{tab:notations}
    \scalebox{0.95}{
    \begin{tabular}{||M{2.1cm}|M{5.8cm}||}
        \hline
        \rowcolor{gray!20}
        Notation & System parameter \\
        \hline
        $b_u^\text{d2d}[t]$ & Indicator of whether UE $u$ is activated in a D2D link in time slot $t$\\
        \hline
        $b_u^\text{track}[t]$ & Indicator of whether UE $u$ is being tracked in time slot $t$\\
        \hline
        $d_u[t]$ & Number of packets delivered to UE $u$ during time slot $t$\\
        \hline
        $d_\text{d2d}[t]$ & Number of packets delivered via D2D link in time slot $t$\\
        \hline
        $d_\text{main}[t]$ & Number of packets delivered via main link in time slot $t$\\
        \hline
        $H_u[t]$ & Blockage condition of UE $u$ at time slot $t$ \\
        \hline
        $I_\text{cb}[t]$ & ID of codebook used in time slot $t$ \\
        \hline
        $I_\text{d2d-rx}[t]$  & UE ID of RX in D2D link \\
        \hline
        $I_\text{d2d-tx}[t]$  & UE ID of TX in D2D link \\
        \hline
        $I_\text{main-dest}[t]$  & UE ID of the destination of packets delivered in main link \\
        \hline
        $I_\text{main-rx}[t]$  & UE ID of RX in main link \\
        \hline
        $I_\text{track}[t]$  & Indicator of whether time slot $t+1$ is a tracking slot \\
        \hline		
        $l_u^\text{block}[t]$ & Number of time slots have passed
        since last observed blockage to UE $u$ \\
        \hline
        $q_u[t]$ & Number of packets in queue of UE $u$ at the beginning of time slot $t$\\
        \hline
        $S^\text{pkg}$ & Fixed size of packet in bit  \\			
        \hline
        $[v_\text{min},v_\text{max}]$ & Speed range of each device \\ 
        \hline
        $[r^\text{rotation}_\text{min},r^\text{rotation}_\text{max}]$ & Rotation rate range of each device \\
        \hline
        $r_u^\text{move}$ & Radius of circular boundary of device $u$\\
        \hline
        $z_u[t]$ & Number of arrived packets for UE $u$ during time slot $t$\\
        \hline
        $N_k^\text{beam}$ & Number of beams in codebook $k$ \\
        \hline
        $N^\text{block}$ & Maximum number of time slots during which a device is under blockage \\
        \hline
        $N^\text{mobility-period}$ & Number of time slots that the mobility pattern of a device got refreshed\\
        \hline
    \end{tabular}\centering
    }	
\end{table}

\subsection{Queue-based network traffic modeling}\label{subsec:netw_queue}
We consider a discrete-time horizon $t=0,1,\hdots$,
where each time step represents a time slot.
The downlink data traffic for UEs arrives at the AP 
in forms of packets with a fixed size in bits, denote as $S^\text{pkg}$.
Denoting $z_u[t]$ as the number of arrived packets for UE $u$ during the $t$-th time slot,
we assume that the packet arrival follows a random process 
$\mathbf{z}[t]=\left[z_1[t],\hdots,z_U[t]\right]^\text{T}$ which is independently and identically distributed (IID) across time slots and its expectation $\mathds{E}\left\{\mathbf{z}[t]\right\}$ is $\boldsymbol{\lambda}=\left[\lambda_1,\hdots,\lambda_U\right]^\text{T}$.
The AP maintains $U$ queues to respectively buffer the packets for the UEs.
Denoting $q_u$ as the number of packets in the queue of UE $u$ at the beginning of the $t$-th time slot,
we define $\mathbf{q}[t]=\left[q_1[t],\hdots,q_U[t]\right]^\text{T}$ as a \textit{queue state vector}. 
With $d_u[t]$ as the number of packets that are successfully delivered to UE $u$ during the $t$-th time slot,
we define $\mathbf{d}[t]=\left[d_1[t],\hdots,d_U[t]\right]^\text{T}$ as the 
\textit{packet departure vector}.
Denoting $\left\{\cdot\right\}^+\triangleq \max(\cdot,0)$, the transition of the state of queues is given as
\begin{equation}
\mathbf{q}[t+1]= \left\{\mathbf{q}[t] - \mathbf{d}[t]\right\}^++ \mathbf{z}[t].
\end{equation}
The stochastic vector $\mathbf{d}[t]$ is determined by the channel quality, user scheduling, and mmWave link configuration, which are quantified in the following subsections.

\subsection{Random mobility and dynamic blockage}\label{subsec:mobility_blockage}
\begin{figure*}[t!]
	\begin{subfigure}[t]{0.35\textwidth}
		\centering
		\includegraphics[width = 0.5\textwidth]{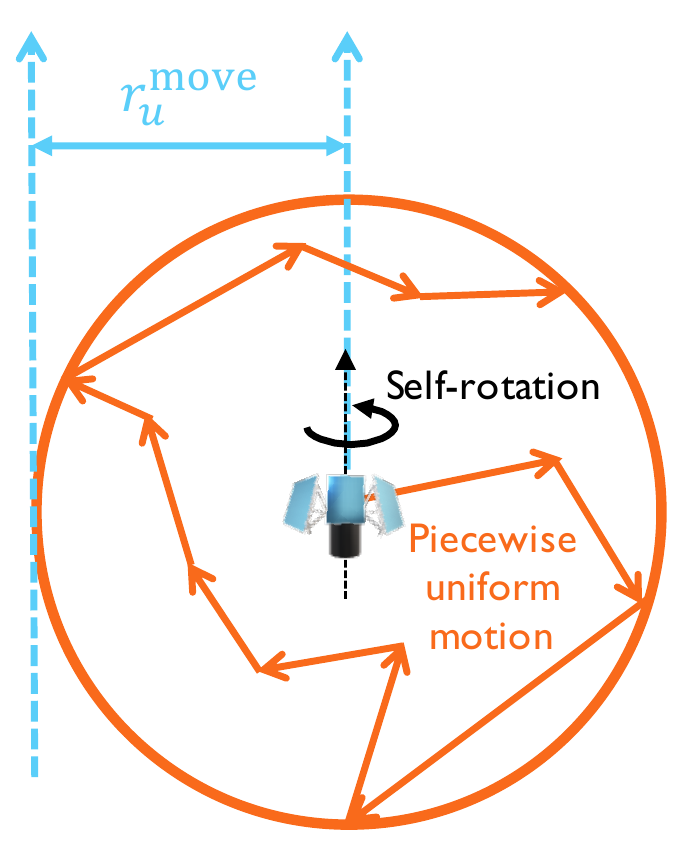}
		\captionsetup{type=figure,justification=centering}
		\caption{Device randomly moves in a region}
		\label{fig:mobility_model}
	\end{subfigure}
	\begin{subfigure}[t]{0.65\textwidth}
		\centering
		\includegraphics[width = 0.6\textwidth]{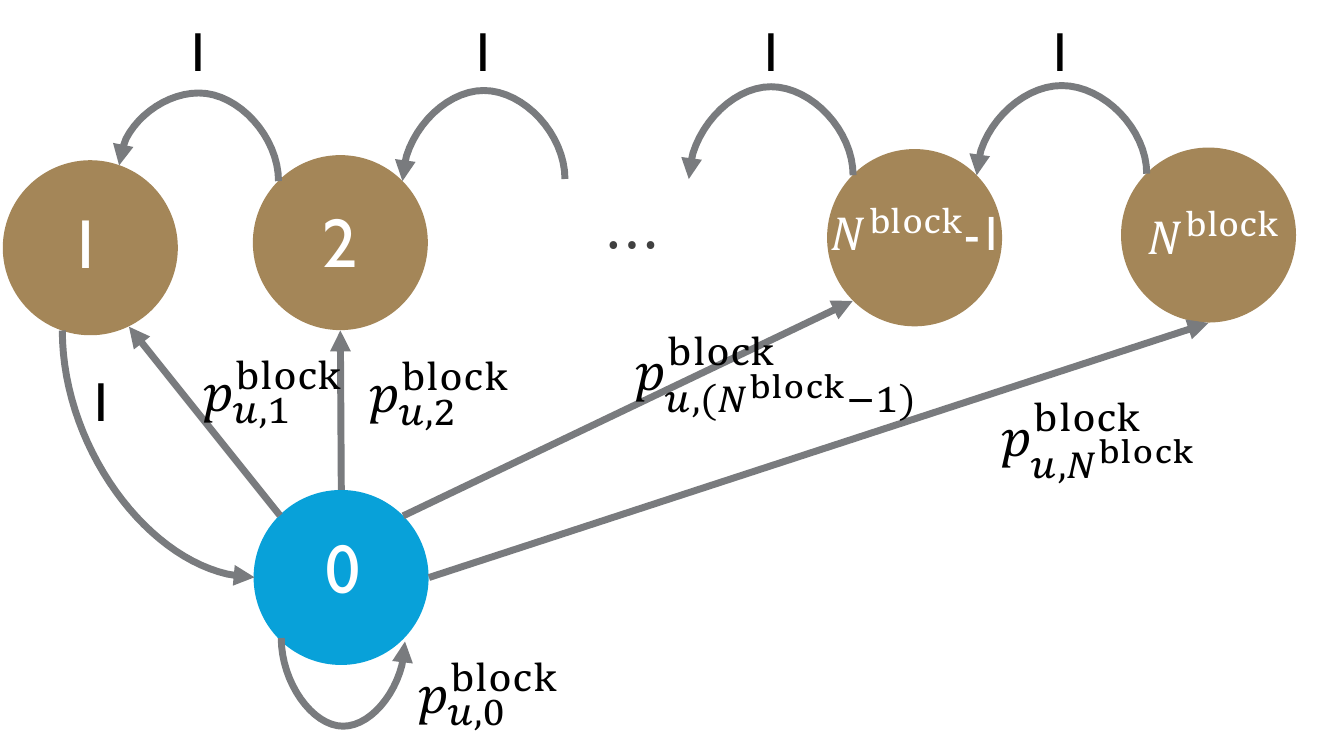}
		\captionsetup{type=figure,justification=centering}
		\caption{Markovian model of the blockage between UE $u$ and AP}
		\label{fig:blockage_model}
	\end{subfigure}
	\caption{Random mobility and dynamic blockage models}
\end{figure*}
\subsubsection{Random mobility}
We assume that each device (UE or AP) randomly moves within its region as shown in Fig.~\ref{fig:mobility_model}, a commonly used model adapted from~\cite{Hse_mobility_model_manet_infocom07}.
We consider that, for every $N^\text{mobility-period}$ time slots, 
each device randomly picks a new speed uniformly from $\left[v_\text{min},v_\text{max}\right]$,
a new movement direction uniformly from $\left[0,2\pi\right]$,
and a new rotation rate uniformly from $\left[r^\text{rotation}_\text{min},r^\text{rotation}_\text{max}\right]$ 
with random direction (clockwise or anti-clockwise along the vertical axis as 2D-movement is considered).
The device then performs a uniform motion with the chosen parameters 
for $N^\text{mobility-period}$ time slots or 
it will reconfigure those parameters when hitting the boundary of its region.
This boundary can be any shape, as we consider in this paper a circle, of which the center is the device's initial position and the radius is parameterized as $r_u^\text{move}$ for device $u$.
\subsubsection{Blockage}
For each UE, we use a $\left(N^\text{block}+1\right)$-state Markov model 
to characterize its blockage condition with respect to the AP.
This is adapted from a two-state blockage model proposed in~\cite{Wang_blockage_wiopt17}.
We denote the blockage condition of UE $u$ at the beginning of the $t$-th time slot as $H_u[t]\in[N^\text{block}]$, which represents that UE $u$ would be under blockage for the following $H_u[t]$ time slots. 
Thus, the UE is under a LOS channel when $H_u[t]=0$ and a NLOS channel when $H_u[t]>0$. 
The state transition is given in Fig.~\ref{fig:blockage_model} where $p_{u,n}^\text{block}$ are unknown.
We assume that the blockage is identified when it happens (e.g. by a drastic change of link quality).
Denoting $l_u^\text{block}[t]$ as the number of time slots have passed
since the last observed blockage to UE $u$, we define \textit{a blockage state vector} $\mathbf{l}^\text{block}[t] = \left[l_1^\text{block}[t],\hdots,l_U^\text{block}[t]\right]^\text{T}$ for later use. 
It is worth pointing out that there could exist a correlation among the blockage conditions of UEs, i.e. $\left\{H_u[t]\right\}_{u=1}^{U}$, given the mobility pattern of the UEs. Since UEs are modeled to be moving with unknown and random direction/velocity/rotation, we assume that $H_u[t]$ are independent among UEs.

\subsection{Packet departure under mmWave link configuration}\label{subsec:link_config}
We now show how the packet departure vector 
$\mathbf{d}[t]$ is determined by the mmWave link configuration.
We first define some notation: 
We denote $b_u^\text{d2d}[t]$ as a binary variable, 
which equals 1 when UE $u$ is activated in a D2D link in the $t$-th time slot. 
We then define $\mathbf{b}^\text{d2d}[t] = \left[b_1^\text{d2d}[t],\hdots,b_U^\text{d2d}[t]\right]^\text{T}$ as a \textit{D2D state vector}.
We denote $I_{\text{d2d-tx}}[t]$ and $I_{\text{d2d-rx}}[t]$ as the ID of the TX and RX in the D2D link respectively, 
when the D2D link exists.
We denote $I_{\text{main-rx}}[t]$ as the ID of the RX in the main link,
and further denote $I_{\text{main-dest}}[t]$ as the ID of the destination of 
the packets transmitted via the main link. 
Accordingly, $I_{\text{main-rx}}[t] \neq I_{\text{main-dest}}[t]$ indicates that a relay is used in the $t$-th slot, 
namely that a D2D link will be activated in the $(t+1)$-th slot. Therefore, the evolution of the D2D state vector can be expressed as 
\begin{align}
b_u^\text{d2d}[t+1] = &\mathds{1}\left\{I_{\text{main-dest}}[t] \neq I_{\text{main-rx}}[t]\right\}\times \nonumber \\
& \mathds{1}\left\{I_{\text{main-dest}}[t] = u \text{~or~} I_{\text{main-rx}}[t] = u\right\}.
\end{align}
We denote $b_u^\text{track}[t]$ as a binary variable, which equals 1
when the $t$-th time slot is a tracking slot for UE $u$.
We then define $\mathbf{b}^\text{track}[t] = \left[b_1^\text{track}[t],\hdots,b_U^\text{track}[t]\right]^\text{T}$ as a \textit{tracking state vector}.
We denote a binary variable $I_{\text{track}}[t]$, 
which equals 1 when the system decides that the time slot $t+1$ is to become a tracking slot.
Therefore, the evolution of the tracking state vector is given as
\begin{equation}
b_u^\text{track}[t+1] = I_\text{track}[t] \mathds{1}\left\{I_\text{main-rx}[t] = u\right\}.
\end{equation}
We denote the ID of the codebook used by the AP for the $t$-th time slot as $I_{\text{cb}}[t]$. 
At the beginning of the $t$-th time slot, the system controller (aka decision maker) uses a policy to make a decision (aka action) on which UE is to be served and how its mmWave link is to be configured. This action is denoted by a quadruple $a[t]=\left( I_{\text{main-dest}}[t], I_{\text{main-rx}}[t], I_{\text{cb}}[t], I_\text{track}[t]\right)$. 
In the following, we will use $a[t]$ to quantify the packet departure vector $\mathbf{d}[t]$.

\subsubsection{Beam training overhead and outage} 
We consider that the $k$-th codebook has $N_k^\text{beam}$ 
non-overlapped directional beams to cover the entire horizontal space (360 degrees), as shown in Fig.~\ref{fig:codebook}. 
We assume that all devices adopt a subarray-based hybrid architecture: The AP and UEs are respectively equipped with 
$N^\text{arr}_\text{AP}$ and $N^\text{arr}_\text{UE}$ phased arrays,
and all the arrays could be simultaneously used to reduce the beam alignment overhead by up to $N^\text{arr}_\text{AP}N^\text{arr}_\text{UE}$ times.
For the $k$-th codebook, we define a \textit{effective coefficient} $C_k^\text{normal}$ which represents the ratio of the time for the DT phase to the entire time slot duration $T^\text{slot}$.
By denoting $T^\text{meas}$ as the time duration of each beam pair testing,
$C_k^\text{normal}$ can expressed as
\begin{equation}\label{eq:coeff_normal}
C_k^\text{normal} = \left(1-\left\lceil\frac{N_k^\text{beam}}{N^\text{arr}_\text{AP}}\right\rceil\left\lceil\frac{N_1^\text{beam}}{N^\text{arr}_\text{UE}}\right\rceil\frac{T^\text{meas}}{T^\text{slot}}\right).
\end{equation}

When beam tracking is to be activated for a time slot, the UE to be scheduled would be the same as the one scheduled in the precedent time slot.
Considering the whole spatial space as a full circle, we perform the beam tracking within a fixed-size circular sector. This circular sector (aka beam tracking region) is equally divided by the direction that is associated with the currently connected beam pair. We then quantize the size of this circular sector in radius and denote it as $\phi^\text{track}$.
We further denote $\mathcal{F}_k^\text{track}$ as the minimum subset of beams in the $k$-th codebook 
that can cover the target beam tracking region. 
Accordingly, we have $\left|\mathcal{F}_k^\text{track}\right|=\left\lceil\frac{\phi^\text{track}}{2\pi}N_k^\text{beam}\right\rceil$.
As a result, the effective coefficient of a tracking slot, denoted by $C_k^\text{track}$, is given as
\begin{equation}\label{eq:coeff_track}
C_k^\text{track} = \left(1-\left|\mathcal{F}_k^\text{track}\right|\left|\mathcal{F}_1^\text{track}\right| \frac{T^\text{meas}}{T^\text{slot}}\right).
\end{equation}
We can see that $C_k^\text{track}\geq C_k^\text{normal}$ as $\frac{\phi^\text{track}}{2\pi}$ is usually smaller than $\frac{1}{N^\text{arr}_\text{AP}}$ and $\frac{1}{N^\text{arr}_\text{UE}}$, which motivates the use of beam tracking when the same UE is served consecutively.

Different codebooks not only result in different overhead reflected by (\ref{eq:coeff_normal}) and (\ref{eq:coeff_track}), but also different levels of robustness to mobility. As shown in Fig.~\ref{fig:outage_model}, narrower beams suffer more from link outages caused by device self-rotation and motion.
To include this in our model, for any link in which the TX has ID $I_\text{tx}$ and uses the $k$-th codebook, and RX has ID $I_\text{rx}$, we define an \textit{outage coefficient} $C_{k,I_\text{tx},I_\text{rx}}^\text{outage}[t]$
which is the ratio of the link outage duration to the entire DT phase.
It is worth pointing out that $C_{k,I_\text{tx},I_\text{rx}}^\text{outage}[t]$ is a random variable as it depends on the transceivers' mobility pattern and channel condition.
As a result, the \textit{eventual effective coefficients} for the main, denoted as $C^\text{eff-main}[t]$ and for D2D link, denoted as $C^\text{eff-d2d}[t]$, can be given as below to incorporate the outage events:
\begin{subequations}
	\begin{align}
	C^\text{eff-main}[t] &= \left(1-C_{I_\text{cb}[t],0,I_\text{main-rx}[t]}^\text{outage}[t]\right) \times \nonumber \\
 &\left[C_{I_\text{cb}[t]}^\text{normal}\left(1-\mathbf{1}^\text{T}\mathbf{b}^\text{track}[t]\right) + C_{I_\text{cb}[t]}^\text{track}\mathbf{1}^\text{T}\mathbf{b}^\text{track}[t]\right],\label{eq:Coeff_main_link}\\
	C^\text{eff-d2d}[t] &= \left(1-C_{1,I_\text{d2d-tx}[t],I_\text{d2d-rx}[t]}^\text{outage}[t]\right)C_1^\text{normal},
	\end{align}	
\end{subequations}
where we recall that
$\mathbf{1}^\text{T}\mathbf{b}^\text{track}[t]=I_\text{track}[t-1]$.

\subsubsection{Effective data rate}
We now formulate the outcome of the BA phase and its impact on the packet departure of the DT phase. 
The following formulation is for a main link and a similar formulation could be done for a D2D link.
Denoting $k$ as the ID of codebook used by the TX, $j_\text{tx}$ and $j_\text{rx}$ are as the ID of beams used by the TX and RX, respectively,
we define a function $f(t,k, I_\text{tx},I_\text{rx},j_\text{tx},j_\text{rx})$ as the RSS obtained between device $I_\text{tx}$ and $I_\text{rx}$.
In particular, $f$ is an unknown channel function that incorporates
the physical layer factors on the received signals, such as previously described UE mobility, blockage condition, 
channel shadowing, hardware impairments, noise and etc.
Denoting $\mathcal{F}[t]$ as the set of beams used by the TX and $\text{rss}_\text{main}[t]$ as the maximum RSS obtained after scanning all the beam pairs, we have
\begin{align}\label{eq:rss_main}
	&\text{rss}_\text{main}[t] = \nonumber \\
 &\max_{\substack{j_\text{tx}\in \mathcal{F}[t], j_\text{rx}\in \left[N^\text{beam}_{1}\right]^+}}
	f\left(t,I_\text{cb}[t],0, I_{\text{main-rx}}[t], j_\text{tx}, j_\text{rx}\right),
\end{align}
where $\mathcal{F}[t] = \left[N^\text{beam}_{I_\text{cb}[t]}\right]^+$ for a normal time slot and $\mathcal{F}[t] = \mathcal{F}_{I_\text{cb}[t]}^\text{track}$ for a tracking time slot.

We consider that there are $M+1$ MCSs that correspond to 
$M+1$ increasing data rates $R_0,\hdots, R_M$.
They are sequentially indexed by MCS $0,\hdots,M$ and MCS 0 (i.e. $R_0=0$) indicates a failure connection. 
Given the RSS in~(\ref{eq:rss_main}), the AP uses the highest supportable MCS for the DT phase by referring to a predefined RSS-MCS table where the minimum RSS required to support MCS $m$ is denoted as $\text{rss}_m$.
Therefore, the instantaneous data rate of the main link is given as
\begin{equation}\label{eq:Ins_R_Main}
R_\text{main}[t] = \max_{m\in[M]} \mathds{1}\left\{\text{rss}_\text{main}[t]\geq \text{rss}_m\right\}R_m.
\end{equation}
It is worth pointing out that we assume that the MCS is selected perfectly and thus there is no error in packet decoding. Though packet retransmission exists in real systems, they could be treated as new arrivals to the queues to fit our model.
Similarly, if a D2D link exists in the $t$-th time slot, its instantaneous data rate can be given as
\begin{subequations}
	\begin{align}
	&R_\text{d2d}[t] = \max_{m\in[M]} \mathds{1}\left\{\text{rss}_\text{d2d}[t]\geq \text{rss}_m\right\}R_m,\\
	&\text{rss}_\text{d2d}[t] = \nonumber\\
 &\max_{j_\text{tx}\in \left[N^\text{beam}_{1}\right]^+, j_\text{rx}\in \left[N^\text{beam}_{1}\right]^+} f\left(t,1,I_{\text{d2d-tx}}[t], I_{\text{d2d-rx}}[t], j_\text{tx}, j_\text{rx}\right) \label{eq:rss_d2d}.
	\end{align}
\end{subequations}
Therefore, the number of packets departed via the main link, denoted as $d_\text{main}[t]$, or via the D2D link, denoted as $d_\text{d2d}[t]$, during the $t$-th slot, are:
\begin{subequations}
	\begin{align}
	&d_\text{main}[t] = \left\lfloor\frac{R_\text{main}[t]C^\text{eff-main}[t]}{S^\text{pkg}}\right\rfloor,\\
	&d_\text{d2d}[t] = \left\lfloor\frac{R_\text{d2d}[t]C^\text{eff-d2d}[t] }{S^\text{pkg}}\right\rfloor.
	\end{align}	
\end{subequations}
We consider that a packet is out of the queue when it has arrived at its destination and 
the maximum number of packets that can be delivered is limited by the capacity of both the main link and the D2D link.
Then the elements of the packet departure vector $\mathbf{d}[t]$ are given as
\begin{equation}
d_u[t] =
\begin{cases}d_\text{main}[t] \mathds{1}\left\{I_{\text{main-dest}}[t] = I_{\text{main-rx}}[t]\right\} &u = I_{\text{main-dest}}[t]\\
\min\left(d_\text{main}[t-1],d_\text{d2d}[t]\right)b_u^\text{d2d}[t] & u = I_{\text{d2d-rx}}[t]\\
0 &\text{otherwise}.
\end{cases}
\end{equation}

Given the formulated system model, our objective is to minimize the packet delay 
by dynamically and jointly scheduling a user for each time slot and configuring its link settings. 
This problem is a complex dynamic decision-making process that involves combinatorial and integer optimization. 
Further, the function $f$ that captures the properties of the wireless system is usually unknown. This makes finding an optimal solution challenging. 
In the following two sections, we will solve it with the aid of RL techniques.

\section{DRL-based joint user and link controller}
In this section, we propose a DRL-based controller to solve this joint user scheduling and link configuration problem.
We first restate the studied problem as a POMDP. Then we show how to adapt the state-of-the-art PPO to our problem.
Finally, we summarize some data scaling operations that are used in training our DRL-based controller.

\subsection{POMDP formulation of joint mmWave UE scheduling and link configuration}\label{subsec:POMDP}
Our studied problem can be characterized as a POMDP described by a 6-tuple $\left(\mathcal{S},\mathcal{D}_0,\mathcal{A},\mathcal{P},\mathcal{R},\gamma\right)$, which is detailed below:
\subsubsection{Observable state space $\mathcal{S}$} 
We define an \textit{observable state} for the $t$-th time slot as a quadruple $s[t] = \left(\mathbf{q}[t],\mathbf{b}^\text{d2d}[t],\mathbf{b}^\text{track}[t],\mathbf{l}^\text{block}[t]\right)$ and the state space $\mathcal{S}$ consists of all possibilities of $s[t]$. There are countably infinite discrete states in $\mathcal{S}$ as $\mathbf{q}[t]$ and $\mathbf{l}^\text{block}[t]$ can be composed of any non-negative integers. We assume the initial state $\mathbf{s}[0]$ follows a distribution $\mathcal{D}_0$.
\subsubsection{Action space $\mathcal{A}$}
The action space $\mathcal{A}$ consists of all the feasible actions that could be potentially taken by a controller.
We recall that the action at the $t$-th time slot has been already defined as the quadruple $a[t] = \left(I_{\text{main-dest}}[t], I_{\text{main-rx}}[t], I_{\text{cb}}[t], I_\text{track}[t]\right)$, where $I_{\text{main-dest}}[t]\in[U]^+$, $I_{\text{main-rx}}[t]\in[U]^+$, $I_{\text{cb}}[t]\in[K]^+$ and $I_\text{track}[t]\in\left\{0,1\right\}$.
It is worth pointing out that $I_\text{track}[t]$ has to be 0 when $I_{\text{main-dest}}[t]\neq I_{\text{main-rx}}[t]$, which is due to the fact that tracking a relay is not allowed in the studied system.
This results in the size of action space $\mathcal{A}$ being calculated as $\left|\mathcal{A}\right|=U^2K + UK$, where $U^2K$ corresponds to the scenario when beam tracking is not activated ($I_\text{track}[t]=0$) while $UK$ corresponds to the other scenario ($I_\text{track}[t]=1$).
\subsubsection{Transition probabilities $\mathcal{P}$}
We denote the probability of being in a state ${\textbf{s}}^\prime$ after taking an action $\mathbf{a}$ in the state ${\textbf{s}}$ 
as $\mathcal{P}\left({s}^\prime,s,a\right)=\mathds{P}\left(s[t+1]={s}^\prime|s[t]=s,a[t]=a\right)$.
Since the mobility and position information of the UEs are not exploited, the underlying transition probabilities between observable states are non-stationary. This makes our studied system a POMDP, where the decision on action is made under the uncertainty and ${\textbf{s}}$ is a partial observation of the true environment.
These transition probabilities $\mathcal{P}$ are unknown since they are decided by the unknown packet arrival process $\mathbf{z}[t]$ and channel function $f$ in (\ref{eq:rss_main}), which jointly reflects the underlying dynamics and randomness of the studied mmWave system (aka environment in RL terminology).
\subsubsection{Reward function $\mathcal{R}$}
In MDP settings, a reward, denoted by $r[t]$ for the $t$-th time slot, is a \textit{random variable} being conditional on the current state $s[t]$ and action taken $a[t]$. Its observation is provided by the environment after the execution of the action, which can be expressed as $r[t]=\mathcal{R}\left(s[t],a[t]\right)$.
In this work, we select $\mathcal{R}$ as the number of packets that are delivered to the destination, i.e. $r[t] = \mathbf{1}^\text{T}\mathbf{d}[t]$.
We will discuss later the motivation of this reward design.

\subsubsection{Discount factor $\gamma$ and related standard definitions in RL settings}
We further present some standard concepts and definitions that are frequently used in RL settings.
With $\gamma\in[0,1]$ representing a discount factor that determines the importance of current and future rewards, the cumulative discounted reward from the $t$-th time slot, denoted by $G[t]$, is defined as~\cite{RL_Intro_Sutton_book18}:
\begin{equation}\label{eq:disc_cum_reward}
	G[t] = \sum\nolimits_{\ell=t}^{\infty}\gamma^{\ell-t} r[l].
\end{equation}
We use $\pi$ to denote a stochastic policy and $\pi\left(a|s\right)$ to denote the probability of choosing an action $a$ when observing a state $s$.
For a policy $\pi$, its \textit{state value function} $v^\pi\left(\cdot\right)$, \textit{state-action value function} $Q^\pi\left(\cdot\right)$, and \textit{advantage function} $A^\pi\left(\cdot\right)$ are defined as below~\cite{A3C_Mnih_ICML16}:
\begin{subequations}\label{eq:rl_v_qv_adv}
	\begin{align}
		&v^\pi\left(s[t]\right) = \mathds{E}_{a[t],s[t+1],a[t+1],\cdots}\left\{G[t]\Large|s[t]\right\},\label{eq:value_func}\\		
		&Q^\pi\left(s[t],a[t]\right) = \mathds{E}_{s[t+1],a[t+1],\cdots}\left\{G[t]\Large|s[t],a[t]\right\},\label{eq:qvalue_func}\\		
		&A^\pi\left(s[t],a[t]\right) = Q^\pi\left(s[t],a[t]\right) - v^\pi\left(s[t]\right).\label{eq:adv_func}
	\end{align}
\end{subequations}
The expectations in (\ref{eq:rl_v_qv_adv}) are taken over the future actions and states.
They are determined by $\pi$ (as $a[t]\sim\pi\left(a[t]|s[t]\right)$) and the unknown transition probabilities $\mathcal{P}$ (as $s[t+1]\sim \mathcal{P}\left(s[t+1],s[t],a[t]\right)$).

By now, we have restated our problem within a POMDP framework using RL terminology.
With our proposed step-wise reward $\mathcal{R}$, i.e. $r[t] = \mathbf{1}^\text{T}\mathbf{d}[t]$, we approximate the original objective of minimizing the average packet delay by maximizing an expected cumulative discounted reward, i.e. $\mathds{E}\left\{\sum_{l=0}^{\infty}\gamma^{l} r[l]\right\}$.
This is motivated by the following facts: 
(1) In queuing networks, minimizing the delay suggests minimizing the sizes of queues as soon as possible according to Little's Law. Therefore, a reward that results in small queues timely would help reduce the mean delay experienced by packets. The queues might however experience a large buffer at the early stage of training/learning, which is not beneficial to NN training as the loss function could have a large variance.
According, we resort to maximizing the total number of packets departed as it is similar to minimizing the queue state given that we have no control over the arrival rates.
(2) In particular, the $\ell_1$-norm of the packet departure vector, i.e. $\mathbf{1}^\text{T}\mathbf{d}[t]$, is an \textit{immediate feedback} of the action taken, which quantifies how much of the queue is immediately reduced, hence is an effective metric for delay minimization. 
(3) Moreover, note that there is a discount factor $\gamma^{l}$ applied to the reward. This implies that the number of packets departed (the step-wise reward) in the immediate future is more important than those in the distant future, which intrinsically meets the objective of minimizing the delay, namely departing more packets as soon as possible.

\subsection{Policy gradient with PPO algorithm within A2C framework}
We now present a DRL-based controller to solve our POMDP problem.
We design this controller by exploiting a state-of-the-art policy gradient method, called the PPO algorithm~\cite{PPO_Schulman_17} within the A2C framework~\cite{A3C_Mnih_ICML16}. 
The major reasons that we use a policy-based method rather than a value-based method are:
(1) Our POMDP has \textit{infinite} discrete states and a large number of actions while it has been empirically shown that DQN~\cite{DQN_Mnih_Nature15} could fail for MDPs with large dimensionality~\cite{PPO_Schulman_17}.
(2) Our POMDP has a \textit{discrete} action space, which prevents us from using DDPG~\cite{DDPG_Lillicrap_ICLR16}.
In the following, we will briefly explain PPO (see~\cite{PPO_Schulman_17} for more details) and show how it is used to train a DRL-based controller in the A2C framework.
\subsubsection{Objective function design with PPO}
\begin{figure}[!t]
	\centering
	\includegraphics[width = 0.45\textwidth]{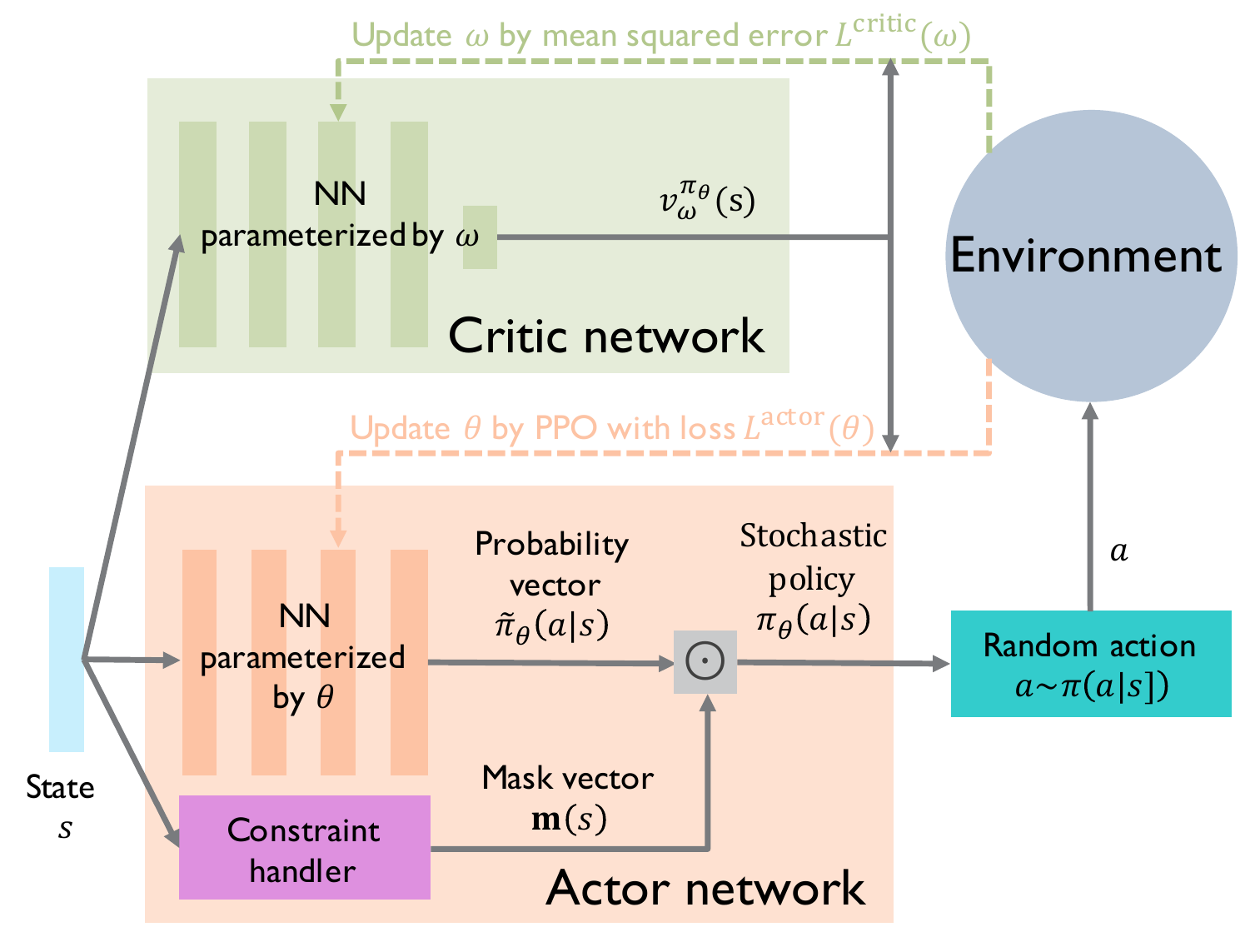}
	\captionsetup{type=figure,justification=centering}
	\caption{A2C framework of proposed DRL-based controller, where $\odot$ is the Hadamard product}
	\label{fig:architecture_drl_agent}
\end{figure}

A policy-based method is to dynamically learn an optimal policy $\pi^*$ that maximizes the objective function $\mathds{E}\left\{\sum_{l=0}^{\infty}\gamma^{l} r[l]\right\}$.
Directly optimizing $\mathds{E}\left\{\sum_{l=0}^{\infty}\gamma^{l} r[l]\right\}$ results in high computational and sample complexity~\cite{PPO_Schulman_17}.
To tackle this, PPO has been proposed in~\cite{PPO_Schulman_17} to focus on the first-order surrogate of the original objective function $\mathds{E}\left\{\sum_{l=0}^{\infty}\gamma^{l} r[l]\right\}$. It was shown to be sample-efficient and have better empirical performance than other policy-based methods, such as vanilla PG~\cite{A3C_Mnih_ICML16} and TRPO~\cite{Schulman_TRPO_ICML15}.
We first present some definitions borrowed from~\cite{PPO_Schulman_17}:
$\pi_{\boldsymbol{\theta}}$ denotes a policy parameterized by ${\boldsymbol{\theta}}$ while ${\boldsymbol{\theta}}_{\text{old}}$ denotes the parameter of the most recently learned policy.
$\hat{A}[t]$ denotes an estimator of the advantage function $A^{\pi_{\boldsymbol{\theta}}}\left(s[t],a[t]\right)$.
$\rho_{\boldsymbol{\theta}} = \frac{\pi_{\boldsymbol{\theta}}\left(a[t]|s[t]\right)}{\pi_{{{\boldsymbol{\theta}}}_{\text{old}}}\left(a[t]|s[t]\right)}$ and $\text{clip}\left(\rho_{\boldsymbol{\theta}},1-\epsilon,1+\epsilon\right)$ is a clipping function that limits the ratio $\rho_{\boldsymbol{\theta}}$ by $1+\epsilon$ when $\hat{A}[t]>0$ and by $1-\epsilon$ when $\hat{A}[t]<0$.
Instead of maximizing $\mathds{E}\left\{\sum_{l=0}^{\infty}\gamma^{l} r[l]\right\}$, PPO opts to maximizes the following surrogate:
\begin{equation}\label{eq:ppo_surrogate}
	L^{\text{clip}}({\boldsymbol{\theta}}) = \mathds{E}\left\{\min\left(\rho_{\boldsymbol{\theta}}\hat{A}[t],\text{clip}\left(\rho_{\boldsymbol{\theta}},1-\epsilon,1+\epsilon\right)\hat{A}[t]\right)\right\}.
\end{equation}
Please refer to~\cite{Schulman_TRPO_ICML15} and~\cite{PPO_Schulman_17} for more details on how (\ref{eq:ppo_surrogate}) is derived from $\mathds{E}\left\{\sum_{l=0}^{\infty}\gamma^{l} r[l]\right\}$. 
With the objective in (\ref{eq:ppo_surrogate}), in the following, we will exploit the A2C framework to learn the estimator $\hat{A}[t]$ and the policy $\pi_{{\boldsymbol{\theta}}}$ via NN, as shown in Fig.~\ref{fig:architecture_drl_agent}.

\subsubsection{DRL-based controller with A2C framework}
As shown in~Fig.~\ref{fig:architecture_drl_agent}, the proposed DRL-based controller has an A2C framework consisting of an actor network and a critic network.
\paragraph{Actor network}
The actor network is composed of a NN parameterized by ${\boldsymbol{\theta}}$ and an additional computational layer which we called \textit{Constraint handler}.
The output of the NN is a probability vector denoted as $\tilde{\pi}_{{\boldsymbol{\theta}}}$. 
The Constraint handler generates a boolean vector which has the same size as $\tilde{\pi}_{{\boldsymbol{\theta}}}$ and is denoted as $\mathbf{m}(s)$.
The Constraint handler sets the elements of $\mathbf{m}(s)$ to either zero or one with the rule: an element is zero if its index corresponds to one of the unfeasible actions that violate the scheduling constraints mentioned in Sec.~\ref{subsec:link_config}.
We recall that the constraints are: 
(1) If a UE is scheduled as either TX or RX in a D2D link, this UE cannot be scheduled as the destination (i.e. RX) in the main link for the same time slot.
(2) If the controller decides, at the $t$-th time slot, to track a UE in the consecutive time slot, then this UE has to be scheduled as the destination (i.e. RX) in the main link in the $(t+1)$-th time slot.
As a result, the output of the actor network is a stochastic policy given as $\pi_{{\boldsymbol{\theta}}}(a\big|s)\sim\tilde{\pi}_{{\boldsymbol{\theta}}}(a\big|s) \odot \mathbf{m}(s)$, where the $\ell_0$-norm of $\pi_{{\boldsymbol{\theta}}}(a\big|s)$ is scaled to 1.
\paragraph{Critic network} 
The critic network is a NN parameterized by ${\boldsymbol{\omega}}$,
which outputs an estimation of the state-value function $v^{\pi_{\boldsymbol{\theta}}}(s)$, denoted by $v^{\pi_{\boldsymbol{\theta}}}_{\boldsymbol{\omega}}(s)$.
This estimation would be used to design $\hat{A}[t]$ with the generalized advantage estimation (GAE) method~\cite{GAE_Schulman_ICLR16}.
In particular, GAE estimates $A^\pi\left(\cdot\right)$ by using the old policy $\pi_{{{\boldsymbol{\theta}}}_{\text{old}}}$ for finite times (aka a batch).
We assume that a batch of $T$ samples is collected before each policy update. 
For notational simplicity, we omit the initial slot index and index these $T$ steps by $\tilde{t} \in [0,T-1]$. GAE proposes to estimate $Q^\pi\left(s[\tilde{t}],a[\tilde{t} ]\right)$ with a linear combination of $T$-step bootstrapping given as below~\cite{GAE_Schulman_ICLR16}:
\begin{align}\label{eq:T_bootstrapping_Q}
&\hat{Q}^\pi\left(s[\tilde{t} ],a[\tilde{t} ]\right) =\nonumber\\
&r[\tilde{t} ] + \gamma r[\tilde{t} +1] +\cdots+ \gamma^{T-\tilde{t} -1}r[T-1] + \gamma^{T-\tilde{t} }v^\pi_{\boldsymbol{\omega}}(s[T]).
\end{align}
Accordingly, an estimator of $A^\pi\left(s[\tilde{t} ],a[\tilde{t} ]\right)$ can be obtained as~\cite{PPO_Schulman_17}:
\begin{subequations}
	\begin{align}
	\hat{A}[\tilde{t}] &= \hat{Q}^\pi\left(s[\tilde{t} ],a[\tilde{t} ]\right) - v^\pi_{\boldsymbol{\omega}}(s[\tilde{t} ]),\\
	& = \delta[\tilde{t} ] + \gamma\delta[\tilde{t} +1] +\cdots+ \gamma^{T-\tilde{t} -1}\delta[T-1],\label{eq:GAE_T_bootstrap}\\
	\delta[\tilde{t} ] & = r[\tilde{t} ] + \gamma v^\pi_{\boldsymbol{\omega}}(s[\tilde{t} +1]) - v^\pi_{\boldsymbol{\omega}}(s[\tilde{t} ]),\label{eq:td_error}
	\end{align}
\end{subequations}
where $\delta[\tilde{t}]$ is called the temporal difference (TD) error in RL literature.

\subsubsection{Loss functions for NN training}
In this part, we present the loss functions used to train the two NNs.
For the actor network, as pointed out by~\cite{A3C_Mnih_ICML16}, an entropy term, denoted by $L^\text{entropy}\left({\boldsymbol{\theta}}\right)$, is added into the loss function to encourage the exploration:
\begin{equation}
L^\text{entropy}\left({\boldsymbol{\theta}}\right) = -\frac{1}{T}\sum_{\tilde{t}=0}^{T}\sum_{a\in\mathcal{A}}\pi_{{\boldsymbol{\theta}}}\left(a\big| s[\tilde{t}]\right) \log\left(\pi_{{\boldsymbol{\theta}}}\left(a\big| s[\tilde{t}]\right)\right).
\end{equation}
An approximation of $L^{\text{clip}}({\boldsymbol{\theta}})$ in (\ref{eq:ppo_surrogate}), denoted as $\hat{L}^{\text{clip}}({\boldsymbol{\theta}})$ can be obtained by taking the expectation over the observed trajectory $\tilde{t}\in [0,T-1]$.
By defining $c_\text{e}$ as a tunable coefficient, the loss function to be minimized for the actor network, denoted by $L\left({\boldsymbol{\theta}}\right)$, is given as
\begin{equation}\label{eq:loss_actor}
L^{\text{actor}}\left({\boldsymbol{\theta}}\right) = - \hat{L}^{\text{clip}}({\boldsymbol{\theta}}) - c_\text{e}L^\text{entropy}\left({\boldsymbol{\theta}}\right),
\end{equation}
For the critic network, the estimation error (i.e. the loss function to be minimized) of $v^{\pi_{\boldsymbol{\theta}}}$ is also approximated with the $T$ samples as below~\cite{PPO_Schulman_17}
\begin{align}\label{eq:loss_critic}
&L^{\text{critic}}({\boldsymbol{\omega}}) = \nonumber\\
&\frac{1}{T}\sum_{\tilde{t}=0}^{T-1}\left(v^{\pi_{\boldsymbol{\theta}}}_{\boldsymbol{\omega}}(s[\tilde{t}]) - \sum_{l=\tilde{t}}^{T-1}\gamma^{l-\tilde{t}}r[l] - \gamma^{T-\tilde{t}}v^{\pi_{\boldsymbol{\theta}}}_{\boldsymbol{\omega}}(s[T-1])\right)^2.
\end{align}
In practical NN implementation, we consider the actor and critic networks as a single united NN (weighted by ${\boldsymbol{\theta}}$ and ${\boldsymbol{\omega}}$) that outputs the $\pi_{{\boldsymbol{\theta}}}(a\big|s)$ and $v^{\pi_{\boldsymbol{\theta}}}_{{\boldsymbol{\omega}}}(s)$ with different blocks. Therefore, the eventual loss function would be the sum of $L^{\text{actor}}\left({\boldsymbol{\theta}}\right)$ and $L^{\text{critic}}({\boldsymbol{\omega}})$.

\subsection{Data scaling for NN training}\label{subsec:data_scalling}
When training a NN, its inputs and outputs are usually scaled (e.g. normalization and standardization) to avoid
unstable learning process or exploding gradients.
Our designed model has involved data scaling in three parts.
(1) We scale the queue state vector $\mathbf{q}[t]$ by dividing it by its maximum element, i.e. $\mathbf{q}[t]\leftarrow\frac{\mathbf{q}[t]}{\left||\mathbf{q}[t]\right||_{\ell_0}}$.
This makes all the elements of $\mathbf{q}[t]$ bounded by 1 while the ratios among the length of queues are still kept.
(2) The elements of $\mathbf{L}^{\text{block}[t]}$ could be large as it is possible that a UE never encounters a blockage. Since ${N}^\text{block}$ is unknown, we introduce a large enough constant $\tilde{N}^\text{block}$ such that $\tilde{N}^\text{block}>{N}^\text{block}$,
and we transform $l_u^{\text{block}[t]}$ as $l_u^{\text{block}[t]} \leftarrow P_u^{\text{block}[t]}\triangleq\max\left(\frac{\tilde{N}^\text{block}-l_u^{\text{block}[t]}}{\tilde{N}^\text{block}+1},0\right)$, which is bounded in $[0,1]$.
$P_u^{\text{block}[t]}$ can be intuitively interpreted as a likelihood that UE $u$ is still undergoing blockage.
This is because when $l^{\text{block}}_{u}$ is large, it is intuitive to guess that the blockage event has already finished.
(3) We scales the reward $r[t]$ as $r[t]\leftarrow \frac{r[t]}{N_\text{p}(x)}$, where $N_\text{p}(x)\triangleq\frac{x 10^9T^\text{slot}}{S^\text{pkg}}$ is the maximum number of packets that could be transmitted when the link rate is $x$ Gbps.
Given the system bandwidth $B$, $x$ could be easily tuned to make the reward $r[t]$ bounded by $[0,1]$.
It is worth pointing out that the data scaling only impacts the NN training but not the algorithm itself.

By now we have illustrated the background theory and implementation of the proposed DRL-based controller, which is summarized in Algorithm~\ref{alg:DRL}.
In this work, we have formulated our POMDP problem in a simplified way. A typical formulation could be done by further specifying the conditional transition probabilities between states, observations, and conditional observation probabilities (please see \cite{ZhuPOMDPs} and reference therein). Besides, the designed NN could further incorporate LSTM component to better approximate the optimal policy.

\begin{algorithm}[!t]
	\caption{Training process for DRL-based controller using PPO within A2C framework}
	\label{alg:DRL}
	\begin{algorithmic}[1]
		\State Input: Total time steps $T^{\text{total}}$, batch size $T$, clipping parameter $\epsilon$, entropy loss coefficient $c_\text{e}$, data scaling coefficient $x$, $\tilde{N}^\text{block}$.
		\State Initialize: randomize NN parameters ${\boldsymbol{\theta}}$ and ${\boldsymbol{\omega}}$; ${\boldsymbol{\theta}}_{\text{old}} \leftarrow {\boldsymbol{\theta}}$; $s[0]\sim \mathcal{D}_0$
		\For {$t=0,\hdots,T^{\text{total}}$}
		\State Scale the state $s[t]$ and input it to the actor/critic networks
		\State Buffer the estimation of the state-value function $v^{\pi_{\boldsymbol{\theta}}}_{\boldsymbol{\omega}}(s[t])$ provided by the critic network
		\State Choose an action $a[t]$ following the stochastic policy $\pi_{{{\boldsymbol{\theta}}}_{\text{old}}}(a|s[t])$ provided by the actor network
		and execute $a[t]$ in system (aka environment)
		\State Get the new state $s[t+1]$ provided by the system
		\State Buffer the scaled reward $r[t]$ provided by the system 
		\If {$ \text{Mod}(t+1,T) == 0$}
		\State reindex the buffered $T$ time steps from $\left\{t,t+1,\cdots,t+T-1\right\}$ to $\left\{0,1,\cdots,T-1\right\}$
		\State Calculate TD error $\delta[\tilde{t}],~\tilde{t}\in[0,T-1]$ as in (\ref{eq:td_error})
		\State Calculate GAE $\hat{A}[\tilde{t}],~\tilde{t}\in[0,T-1]$ as in (\ref{eq:GAE_T_bootstrap})
		\State Optimize the NN parameter ${\boldsymbol{\theta}}$ (actor network) by $\nabla_{\boldsymbol{\theta}} L^{\text{actor}}({\boldsymbol{\theta}})$ as in (\ref{eq:loss_actor})
		\State Optimize the NN parameter ${\boldsymbol{\omega}}$ (critic network) by $\nabla_{\boldsymbol{\omega}} L^{\text{critic}}({\boldsymbol{\omega}})$ as in (\ref{eq:loss_critic})
		\State $\pi_{{{\boldsymbol{\theta}}}_{\text{old}}} \leftarrow \pi_{\boldsymbol{\theta}}$
		\State Clear the buffer
		\EndIf
		\EndFor
	\end{algorithmic}
\end{algorithm}

\section{Empirical MAB-based joint user and link controller}
DRL-based solutions to MDP problems generally result in a high sample and computational complexity due to NN training. This could be a huge challenge in wireless applications, where feedbacks are usually online available unless a well-modeled simulator could be built.
Motivated by this, in this section, we further propose another learning-based controller which is sample-efficient and has low computational complexity.
It is called the \textit{Empirical MAB-based controller}. Its key idea is to decompose the decision-making at each time slot into a sequence of sub-actions as follows: First, an empirical maxweight policy is exploited for UE scheduling.
Thereafter, the MAB framework, one of the most basic RL settings, is leveraged to learn whether and which relay should be configured as the main link.
Then the MAB framework is again used to learn the optimal codebook for the configured main link.
Finally, a heuristic rule is derived to decide whether beam tracking is performed, which completes the whole decision-making process.

\subsection{UE scheduling with empirical maxweight}\label{subsec:max_weight}
It is well-known that the queue-aware maxweight scheduling policy is throughput optimal~\cite{MaxWeight_TIT_93}.
Given the instantaneous packet departure rate of UE $u$ at time slot $t$ for $u\in[U]^+$, denoted as $\tilde{d}_u[t]$, the maxweight allocates time slot $t$ to the user that satisfies
\begin{equation}\label{eq:max_weight}
	\tilde{u} = \arg\max_{u\in[U]^+} q_u[t]\tilde{d}_u[t].
\end{equation}
Unfortunately, $\tilde{d}_u[t]$ is unknown in our studied system due to the fact that $\tilde{d}_u[t]$ is the function of random channel and stochastic link configuration.
Therefore, we propose to use the empirical mean of $\tilde{d}_u[t]$ to perform the UE selection described in (\ref{eq:max_weight}).
To be specific, a sample of $\tilde{d}_u[t]$ is observable at the end of time slot $t$ when UE $u$ is activated as the RX in either a main or D2D link, which is captured by the packet departure vector $\mathbf{d}[t]$. 
We denote $N^\text{rx}_u[t]$ as the number of times that UE $u$ has been the RX in any link until the beginning of time slot $t$, then we have
\begin{equation}\label{eq:counter_update}
	N^\text{rx}_u[t] = N^\text{rx}_u[t-1]+\mathds{1}\left\{u=I_{\text{main-rx}}[t] \text{~or~} u=I_{\text{main-dest}}[t-1]\right\}.
\end{equation}
We then propose an empirical estimation of $\tilde{d}_u[t+1]$, denoted by $\hat{d}_u[t+1]$, given in (\ref{eq:empi_mean_Dt}), where the coefficient $\frac{1}{2}$ in  $\frac{1}{2}\mathds{1}\left\{B^\text{d2d}_u[t]=1\right\}$ is to penalize the extra delay (one time slot) that is introduced by the use of a D2D link.
As a result, our empirical maxweight policy for the UE scheduling can be summarized in (\ref{eq:max_weight_ue}), where the first case deals with scenarios when beam tracking is activated while the second one tackles the constraint that UEs in an activated D2D link could not be scheduled for a main link. 
\begin{figure*}[!t]
\normalsize
\begin{equation}\label{eq:empi_mean_Dt}
	\hat{d}_u[t+1] = \frac{N^\text{rx}_u[t-1]\hat{d}[t]+d_u[t]\left(\frac{1}{2}\mathds{1}\left\{b^\text{d2d}_u[t]=1\right\} + \mathds{1}\left\{u=I_\text{main-rx}[t]\right\} \right)    }{N^\text{rx}_u[t]},
\end{equation}
\hrulefill
\vspace*{-6pt}
\end{figure*}
\begin{figure*}[!t]
\normalsize
\begin{equation}\label{eq:max_weight_ue}
I_{\text{main-dest}}[t] =
\left\{
\begin{array}{lr}
I_{\text{main-dest}}[t-1], & \mathbf{1}^\text{T}\mathbf{b}^\text{track}[t] = 1,\\
\arg\max_{u\in\left\{i\in [U]^+ \big|B_i^\text{d2d}[t]=0\right\}} q_u[t]\hat{d}_u[t], &  \text{otherwise},
\end{array}
\right.
\end{equation}
\hrulefill
\vspace*{-6pt}
\end{figure*}

\subsection{MAB-based relay and codebook selection for UE $I_{\text{main-dest}}[t]$}
\subsubsection{MAB-based relay selection}\label{subsubsec:mts_relay}
\paragraph{Viewing relay selection from perspective of MAB} 
Without considering the state information $s[t]$, selecting a relay (could be any UE in the network) for UE $I_\text{main-dest}[t]$ can be regarded as a MAB problem, in which UE $u$ is called arm $u$ and a reward of the arm is provided each time when it is used. 
The arms are dynamically explored and exploited as time evolves such that the expected cumulative reward is maximized.
As the state $s[t]$ is omitted, a MAB problem is also called single-state (or stateless) RL problem~\cite{RL_Intro_Sutton_book18}.
At each time slot, one of the $U$ arms (relays) is chosen to be a potential relay for the scheduled UE $I_{\text{main-dest}}[t]$, and its network ID is denoted as $I_\text{main-rx}[t]$.
The case $I_\text{main-rx}[t]=I_\text{main-dest}[t]$ simply represents that using UE $I_\text{main-dest}[t]$ itself as a 'relay', namely that no D2D transmission would be used in the next time slot. 

\paragraph{Designing arm reward}
With the chosen relay $I_\text{main-rx}[t]$, the instantaneous data rate $R_\text{main}[t]$ (in (\ref{eq:Ins_R_Main})) is observed by completing the BA phase and the effective coefficient $C^\text{eff-main}[t]$ (in (\ref{eq:Coeff_main_link})) is observed at the end of the DT phase.
Accordingly, we design the reward of arm $u$ as the eventual effective data rate provided by relay $u$, which is expressed in (\ref{eq:reward_relay_selection}), where the coefficient $\frac{1}{2}$ in the second case is to penalize the extra delay introduced by the D2D link (similar to that in (\ref{eq:empi_mean_Dt})). 
\begin{figure*}[!t]
\normalsize
\begin{equation}\label{eq:reward_relay_selection}
R_{\text{relay},u}[t] =
\left\{
\begin{array}{lr}
R_{\text{main}}[t]C^\text{eff-main}[t], & u = I_\text{main-rx}[t],\\
\frac{1}{2}\text{min}\left(R_{\text{main}}[t]C^\text{eff-main}[t],R_{\text{d2d}}[t+1]{C}^\text{d2d}[t+1]\right), &  \text{otherwise},
\end{array}
\right.
\end{equation}
\hrulefill
\vspace*{-6pt}
\end{figure*}
The motivation for using the eventual effective data rate in the reward is to penalize the uses of bad relays that provide poor link quality (e.g. UEs that are under blockage or too far away UE $I_\text{main-dest}[t]$).
It is worth pointing out that $R_{\text{relay},u}[t]$ in (\ref{eq:empi_mean_Dt}) would be a delayed reward when $u \neq I_\text{main-rx}[t]$.

\paragraph{Selecting relay by Thompson sampling (TS)-based algorithm}
TS is well-known for its simple implementation and outstanding empirical performance~\cite{Empirical_TS_Chapelle_NIPS2011} in solving MAB problems.
The essence of TS is, at each time step, to choose an arm by sampling the predefined prior distributions of all arms' rewards.
Then the reward provided by executing the chosen action is further used to update the posterior distributions of those priors.
As this process repeats, the posteriors converge to the true distributions of the arms' rewards.
We briefly introduce its adaptation in this paper by detailing the prior distribution, sampling rule, and posterior update.
Please refer to our prior work~\cite{TSCB_YiZhang_Mobihoc_2021} for more details.

\textbf{Prior distribution of arm reward}:
As there are $M+1$ MCS levels, the instantaneous data rate provided by arm $u$, i.e., $R_{\text{main}}[t]$ or $R_{\text{d2d}}[t+1]$, actually follows a \textit{one-trial multinomial distribution} with  support $\left\{R_0,\cdots,R_M\right\}$ and an unknown probability vector $\mathbf{P}^\text{relay}_u[t] = \left[P^\text{relay}_{0,u}[t],\cdots,P^\text{relay}_{M,u}[t]\right]^\text{T}$, where $P^\text{relay}_{m,u}[t]$ is the probability of using MCS $m$ at time slot t when using arm $u$.
Therefore, the conjugate prior of a multinomial distribution, i.e. a Dirichlet distribution, is a perfect prior for estimating $\left\{\mathbf{P}^\text{relay}_u[t]\right\}_{u=1}^U$.
To fully characterize the true reward of arm, i.e. effective data rate $R_{\text{relay},u}[t]$, the effect coefficients in~(\ref{eq:reward_relay_selection}) should be considered. We will handle this in the posterior update.

\textbf{Sampling rule}:
At the beginning of each time slot, the controller samples the priors of the available arms (i.e. the UEs which are not in a D2D link) and chooses the arm that provides the largest predictive reward, as described by Line 6-9 in Algorithm~\ref{alg:HMAB}.

\textbf{Posterior update}:
Note that $R_{\text{relay},u}[t]$ is not a timely reward when a D2D link is to be activated. Accordingly, the posterior update would be processed whenever the reward is available.
We take the case with a timely reward for example, which is described by Line 18-22 in Algorithm~\ref{alg:HMAB}: 
since $C^\text{eff-main}[t]$ is revealed when time slot $t$ ends, we randomize the $R_{\text{main},u}[t]$ by generating a Bernoulli random variable $X_\text{relay}\sim \text{Bernoulli}\left(\tilde{C}^\text{eff-main}[t]\right)$, and forcing $R_{\text{relay},u}[t]=0$ when $X_\text{relay}=0$.
This randomization incorporates the effect of $C^\text{eff-main}[t]$ into $\mathbf{P}^\text{relay}_u[t]$, hence helping characterize the true reward of an arm in form of a multinomial distribution.
For the case with delayed reward, a similar process is performed as shown by Line 23-34 Algorithm~\ref{alg:HMAB}.

\subsubsection{MAB-based codebook selection}\label{subsubsec:mts_bw}
Similar to relay selection, the codebook selection could be regarded as a MAB problem with $K$ arms ($K$ codebooks at the AP).
The reward of the $k$-th codebook is $R_{\text{cb},k}[t] = R_\text{main}[t]C^\text{eff-main}[t]~\left(\text{with}~ I_\text{CB[t]}= k\right)$.
It is worth pointing out that this is an optimally designed reward since the objective of codebook selection is to find out the optimal beamwidth for the main link, which is regardless of user scheduling and relay selection.
A similar TS-based algorithm could be applied to solve the codebook selection problem. 

Two remarks on the application of TS in relay and codebook selection are provided below:
\begin{remark}\label{remark_1}
	It has been shown in our prior work~\cite{TSCB_YiZhang_Mobihoc_2021} that the TS-based algorithm used in Sec.~\ref{subsubsec:mts_relay} and Sec.~\ref{subsubsec:mts_bw} is asymptotically optimal in choosing the best arm when the rewards are IID across the time slots.
	This condition may hold for codebook selection but not for the relay selection.
	This is because the reward distributions of the relays are time-variant due to the learning process of the codebook selection and beam tracking.
	Nevertheless, the convergence of the learning process is confirmed by our later evaluations.
\end{remark}
\begin{remark}\label{remark_2}
	We did not merge the $U$-armed relay selection and the $K$-armed codebook selection into a single one with $UK$ arms due to the fact that they have different reward functions.
\end{remark}

\subsection{Heuristic beam tracking}\label{subsec:heu_track}
We finally derive a decision rule on whether to conduct beam tracking.
It is designed based on the following observation: if the same UE is to be chosen by (\ref{eq:max_weight_ue}) for the consecutive time slot,
it is definitely beneficial to activate the beam tracking mechanism.
Accordingly, we empirically predict the sizes of queues as
\begin{equation}
\hat{q}_u[t+1] = \left\{q_u[t] - \hat{d}_{I_{\text{main-rx}}[t]}[t]\mathds{1}\left\{u =I_{\text{main-rx}}[t]\right\}\right\}^+ + \hat{Z}_u[t],
\end{equation}
where $\hat{Z}_u[t] = \frac{(t-1)\hat{Z}_u[t-1]+ Z_u[t]}{t}$ is the empirical mean of the packet arrival rate of UE $u$. Therefore, the UE to be served in time slot $t+1$, denoted by $\hat{I}_{\text{main-rx}}[t+1]$, can be predicted using (\ref{eq:max_weight_ue}) as
\begin{equation}
	\hat{I}_{\text{main-rx}}[t+1] = \arg\max_{u\in[U]^+} \hat{q}_u[t+1]\hat{d}_u[t].
\end{equation}
As a result, a heuristic decision rule on whether beam tracking is to be performed is given as
\begin{equation}\label{eq:tracking_decision}
I_{\text{track}}[t]  =
\left\{
\begin{array}{lr}
0,~~~~~~~~~~~~~~~~~~~~~~~I_{\text{main-dest}}[t] \neq I_{\text{main-rx}}[t],\\
\mathds{1}\left\{\hat{I}_{\text{main-rx}}[t+1] = I_{\text{main-rx}}[t] \right\},~~~~~\text{otherwise}.
\end{array}
\right.
\end{equation}

By now, we have illustrated our proposed Empirical MAB-based controller and its whole process is summarized in Algorithm~\ref{alg:HMAB}.
In particular, we initialize the algorithm by exploring each feasible action $a[t]=\left( I_{\text{main-dest}}[t], I_{\text{main-rx}}[t], I_{\text{cb}}[t], I_\text{track}[t]\right)$ for at least $N_\text{EMAB-Init}$ times to reach a good initial state of the priors and those empirical means.
It is worth pointing out that this Empirical MAB-based controller is an approximation and combination of several classic algorithms which have provable performance guarantees.
Providing a reward (or regret) analysis to this empirical combination is however challenging due to that the TS is applied twice sequentially and that the reward distribution of relay selection is time-variant due to the user scheduling, which is pointed out in the Remarks~\ref{remark_1} and~\ref{remark_2}. Hence a theoretical analysis of the proposal empirical algorithm is out of the scope of this work.
We will confirm the feasibility of the proposed learning controller by numerical evaluation in Sec.~\ref{sec:evaluation}.

\begin{algorithm}
	\caption{Empirical MAB-based controller (To be continued)}
	\label{alg:HMAB}
	\begin{algorithmic}[1]
		\State Input: Total time steps $T^{\text{total}}$; number of UE $U$, number of codebooks $K$; number of non-zero MCSs $M$; rate vector $\mathbf{r}=\left[r_0,r_1,\hdots,r_M\right]^{\text{T}}$; initial exploration number $N_\text{EMAB-Init}$.
		\State Initialize: $s[0]\sim \mathcal{D}_0$; $\alpha^\text{relay}_{m,u,u} = 1, m\in[M], u \in [U]^+$; $\alpha^\text{cb}_{m,k,u} = 1, m\in[M],k \in [K]^+, u\in[U]^+$; $\hat{\mathbf{d}}[0] = \mathbf{0}$; $\hat{\mathbf{q}}[0] = \mathbf{0}$; $\hat{\mathbf{z}}[0] = \mathbf{0}$; counters for empirical means.
		\State Initialize: Use each feasible action for at least $N_\text{EMAB-Init}$ times and update priors, empirical means and counters accordingly.
		\For {$t=N_\text{EMAB-Init} |\mathcal{A}|,\hdots,T^{\text{total}}$}
			\State Decide $I_{\text{main-dest}}[t]$ as in (\ref{eq:max_weight_ue}). \Comment{Empirical maxweight UE scheduling}
			\For {$u\in \left\{i\in [U]^+ \big|B_i^\text{d2d}[t]=0\right\}$}
				\State Sample $\mathbf{d}_u\sim Dir\left(\alpha_{0,u,I_{\text{main-dest}}[t]}^\text{relay},\hdots,\alpha_{M,u,I_{\text{main-dest}}[t]}^\text{relay}\right)$.
			\EndFor
			\State $I_\text{main-rx}[t] = \arg\max_{u\in\left\{i\in [U]^+ \big|B_i^\text{d2d}[t]=0\right\}} \mathbf{r}^{\text{T}} \mathbf{d}_u$. \Comment{MAB-based relay selection}
			\For {$k\in[K]^+$}
				\State Sample $\mathbf{d}_k\sim Dir\left({\alpha}_{0,k,I_{\text{main-rx}}[t]}^\text{cb},\hdots,{\alpha}_{M,k,I_{\text{main-rx}}[t]}^\text{cb}\right)$.
			\EndFor
			\State $I_\text{cb}[t] = \arg\max_{k\in[K]^+} \mathbf{r}^{\text{T}} \mathbf{d}_k$. 
			\Comment{MAB-based codebook selection}
			\State Decide $I_{\text{track}}[t]$ as in (\ref{eq:tracking_decision}). 
			\Comment{Heuristic beam tracking}
		\State Execute action $a[t] = \left(I_{\text{main-dest}}[t], I_{\text{main-rx}}[t], I_{\text{cb}}[t], I_\text{track}[t]\right)$ in system.
		\State Lookup the RSS-MCS table to get $R_\text{main}[t]=R_{m_\text{main}[t]}$ with $m_\text{main}[t]\in [M]$.
		\State Observe the effective coefficient $C^\text{eff-main}[t]$ at the end of the DT phase.
		\If {$I_\text{main-dest}[t]=I_\text{main-rx}[t]$}
			\Comment{Prior update with timely reward of relay selection}
			\State Generate a Bernoulli random variable $X_\text{relay}\sim \text{Bernoulli}\left(\tilde{C}^\text{eff-main}[t]\right)$.
			\State $m_\text{relay}[t] = m[t] X_\text{relay}$.
			\State Prior update: $\alpha^\text{relay}_{m_\text{relay}[t],I_{\text{main-rx}}[t],I_{\text{main-dest}}[t]}:=\alpha^\text{relay}_{m_\text{relay}[t],I_{\text{main-rx}}[t],I_{\text{main-dest}}[t]}+1$.
		\EndIf
		\If{$\mathbf{1}^\text{T}\mathbf{b}^\text{d2d}[t]\neq 0$}
			\Comment{Prior update with delayed reward of relay selection} 
			\State Lookup the RSS-MCS table to get $R_\text{d2d}[t]=R_{m_\text{d2d}[t]}$ with $m_\text{d2d}[t]\in [M]$.
			\State Observe the effective coefficient $C^\text{eff-d2d}[t]$ at the end of the DT phase.
			\If {$R_\text{d2d}[t] < R_\text{main}[t-1]$}
				\State Generate a Bernoulli random variable $X_\text{relay}\sim \text{Bernoulli}\left(\frac{1}{2}C^\text{eff-main}[t-1]\right)$.
				\State $m_\text{Relay-D2D} = m_\text{main}[t-1] X_\text{relay}$.
			\Else
                    \State Generate a Bernoulli random variable $X_\text{relay}\sim \text{Bernoulli}\left(\frac{1}{2}C^\text{eff-d2d}[t]\right)$.
				\State $m_\text{Relay-D2D} = m_\text{d2d}[t] X_\text{relay}$.
			\EndIf
				\algstore{part1}
				\end{algorithmic}
				\end{algorithm}
				\setcounter{algorithm}{1}
				\begin{algorithm}
				\caption{Empirical MAB-based controller (Continued) }
				\begin{algorithmic}[1]
				\algrestore{part1}
			\State Prior update: $\alpha^\text{relay}_{m_\text{Relay-D2D},I_{\text{d2d-tx}}[t],I_{\text{d2d-rx}}[t]}:=\alpha^\text{relay}_{m_\text{Relay-D2D},I_{\text{d2d-tx}}[t],I_{\text{d2d-rx}}[t]}+1$.
		\EndIf
		\State Generate a Bernoulli random variable $X_\text{cb}\sim \text{Bernoulli}\left(C^\text{eff-main}[t]\right)$.
		\Comment{Prior update with timely reward of codebook selection} 
		\State $m_\text{cb}[t] = m[t] X_\text{cb}$. 
		\State Prior update: $\alpha^\text{cb}_{m_\text{cb}[1],I_{\text{cb}}[t],I_{\text{main-rx}}[t]}:=\alpha^\text{cb}_{m_\text{cb}[t],I_{\text{cb}}[t],I_{\text{main-rx}}[t]}+1$.
		\State Update empirical means $\hat{\mathbf{d}}[t+1]$, $\hat{\mathbf{q}}[t+1]$, $\hat{\mathbf{z}}[t+1]$, and corresponding counters.
		\EndFor
	\end{algorithmic}
\end{algorithm}

\section{Evaluation results}\label{sec:evaluation}
In this section, we provide the performance evaluation of the proposed two RL-based controllers.
We first illustrate the system and training setups in Sec.~\ref{sub:sim_sys_setup} and Sec.~\ref{sub:sim_train_setup}, respectively.
Then we show the performance comparison in Sec.~\ref{sub:sim_per_comp}. 

\subsection{System setup}\label{sub:sim_sys_setup}
We simulate an outdoor mmWave system that operates at a center frequency $f_c=60$ GHz with a total bandwidth $B=2.16$ GHz~\cite{802.11ad_standard:2016}.
Since the outdoor mmWave channel is generally sparse and codebook-based beam alignment is performed, the LOS path could be well estimated and dominant if it exists.
We only consider the RSS of a dominant path after the BA phase.
The path loss of the LOS path between any two devices can be modeled as~\cite{5G_NR_standard_V16:2019}
\begin{equation}\label{eq:LOS_PL}
	PL^\text{LOS} \text{(dB)} = 28+22\log_{10}(d)+20\log_{10}(f_c)+\mathcal{X},
\end{equation}
where $d$ is the distance between the transceivers and $\mathcal{X}$ is the shadowing fading that follows the normal distribution $\mathcal{N}(0,\sigma^2)$.
Generally, a NLOS path could suffer from more than 15 dB path loss (or even worse) than a LOS path loss~\cite{5G_NR_standard_V16:2019}.
Accordingly, we assume an additional path loss $PL^\text{block}$ would be added to $PL^\text{LOS}$
when the LOS link does not exist, which yields:
\begin{equation}\label{eq:NLOS_PL}
	PL^\text{NLOS} \text{(dB)} = PL^\text{LOS} + PL^\text{block}.
\end{equation}
In this work, when a blockage event happens to a certain link (AP-to-UE or UE-to-UE link), we model its blockage loss $PL^\text{block}$ as a random variable that is uniformly distributed between 10 to 30 dB. When the random variable $PL^\text{block}$ has a relatively small value, it could be interpreted as the path loss of the NLOS link provided by the potential reflectors or scatterers. When $PL^\text{block}$ has a relatively larger value, it can be viewed as an outage happens.
As we can see from $(\ref{eq:LOS_PL})$ and $(\ref{eq:NLOS_PL})$, the channel fluctuation in the simulation, 
collectively referred to as the path loss represented by $PL$,
is caused by the device mobility, link blockage, and shadowing fading.
We denote $W$ as the link marginal budget and implementation loss, $P_\text{T}$ is the transmitting power,
which is set as $P_\text{T}\triangleq P_\text{AP}=15$ dBm for the AP (main link) 
and $P_\text{T}\triangleq P_\text{AP}=10$ dBm for UEs (D2D link).
We use $G_\text{T}~(\text{or}~G_\text{R})\triangleq\frac{16\pi}{6.67b^\text{azi}b^\text{ele}}$~\cite{MiWEBA} 
to represent the antenna gain of a beam pattern whose azimuth width is $b^\text{azi}$ and elevation width is $b^\text{ele}$.
We recall that only $b^\text{azi}$ is configurable via the codebook selection, i.e. $b^\text{azi}(k)=\frac{2\pi}{N^\text{beam}_k}$.
With the above notation, the RSS obtained by a link after beam alignment can be given as
\begin{equation}\label{eq:RSS_sim}
	RSS \text{(dB)} = P_\text{T} + G_\text{T} + G_\text{R} - PL -W.
\end{equation}
The associated SNR of the link could be further calculated as $SNR = RSS - P_\text{N}$, where $P_\text{N}$ is the bandwidth-dependent noise power and $P_\text{N} = -174 + 10\log_{10}B + 10$.
For convenience, we summarize all relevant simulation parameters and their values in Table~\ref{tab:sys_paras}.
\begin{table}[t!]
    \renewcommand*{\arraystretch}{1.2}
    \captionsetup{type=table,justification=centering}
    \captionof{table}[t]{Simulation parameters} 
    \label{tab:sys_paras}
    \scalebox{0.95}{
    \begin{tabular}{||M{6.0cm}|M{2.0cm}||}
    \hline
    \rowcolor{gray!20}
    System parameter & Value \\
    \hline
    Carrier frequency ($f_c$) & 60 GHz \\
    \hline
    Duration of a time slot ($T^\text{slot}$) & 10 ms\\
    \hline
    Duration of a beam pair testing ($T^\text{meas}$) & 10 $\mu$s\\
    \hline
    Size of tracking region ($\phi^\text{track}$) & $\frac{\pi}{6}$ ($30^\circ$)\\
    \hline
    Packet size ($S^\text{pkg}$) & 2312$\times$8 bits \\
    \hline
    Total downlink data traffic ($\sum_{u=1}^{U}\lambda_u S^\text{pkg}$) & 1 Gbps \\
    \hline
    Shadowing fading variance ($\sigma$) & 2 dB\\
    \hline
    Noise power ($P_\text{N}$) & -70.655 dBm\\
    \hline
    Link budget/implementation loss ($W$) & 10 dB\\
    \hline
    \rowcolor{gray!20}
    Hardware parameter & Value \\
    \hline
    Number of arrays of AP ($N^{\text{arr}}_\text{AP}$) & 4 \\		
    \hline
    Number of arrays of UE ($N^{\text{arr}}_\text{UE}$) & 4 \\		
    \hline
    Number of codebooks at AP ($K$) & 6 \\
    \hline
    Number of beams in $K$ codebooks ($N_k^\text{beam}, k\in[K]^+$) & {24, 32, 64, 128, 256, 512} \\
    \hline
    Elevation beamwidth $b^\text{ele}$ & $75^\circ$ \\
    \hline
    Transmitting power of AP ($P_\text{AP}$) & 15 dBm \\
    \hline
    Transmitting power of UE ($P_\text{UE}$) & 10 dBm \\
    \hline
    \rowcolor{gray!20}
    Blockage and mobility parameter & Value \\
    \hline
    Blockage attenuation $PL^\text{block}$ & [10, 30] dB\\
    \hline
    Minimum slots in blockage  & 2 \\
    \hline
    Maximum slots in blockage ($N^\text{block}$) & 6 \\
    \hline
    Speed range $[v_\text{min},v_\text{max}]$ & [0,10] m/s \\
    \hline
    Rotation rate $[r^\text{rotation}_\text{min},r^\text{rotation}_\text{max}]$ & [0,10] degree/s \\
    \hline
    Mobility update periodicity ($N^\text{mobility-period}$) & $20$\\
    \hline
    Radius of circular boundary of Device ($r_u^\text{move}$) & 5 m\\
    \hline
\end{tabular}\centering
    }
\end{table}

\subsection{Training parameters and setups}\label{sub:sim_train_setup}
\subsubsection{Relevant parameters}
Both the actor and critic networks of the DRL-based controller have three hidden dense layers and each layer has 128 units. We use the Adam optimizer to update the NN parameters with a learning rate $\ell_\text{r} =0.001$. In particular, $\ell_\text{r}$ decays every 20 updates with a decay coefficient 0.9. The clipping parameter $\epsilon$ is 0.2 and the discounting factor $\gamma$ is 0.999. The entropy loss coefficient $c_\text{e}$ is 0.05. The data scaling coefficients are set as $x=2$, $\tilde{N}^\text{block}=10$.
For the Empirical MAB-based controller, the initial exploration number $N_\text{EMAB-Init}$ is set to 5.
\subsubsection{Training setups}
We monitor the learning process of our proposed solutions for 240 iterations, where one iteration consists of 1500 slots, 
which corresponds to 15 seconds as $T^\text{slot} = 10$ ms. 
The total monitored time steps $T^{\text{total}}$ is 36,000 seconds, i.e one hour.
Unless otherwise stated, the batch size $T$ is 5, namely that the NNs are updated every 5 time slots. 
\begin{figure}
		\centering
		\includegraphics[width = 0.5\textwidth]{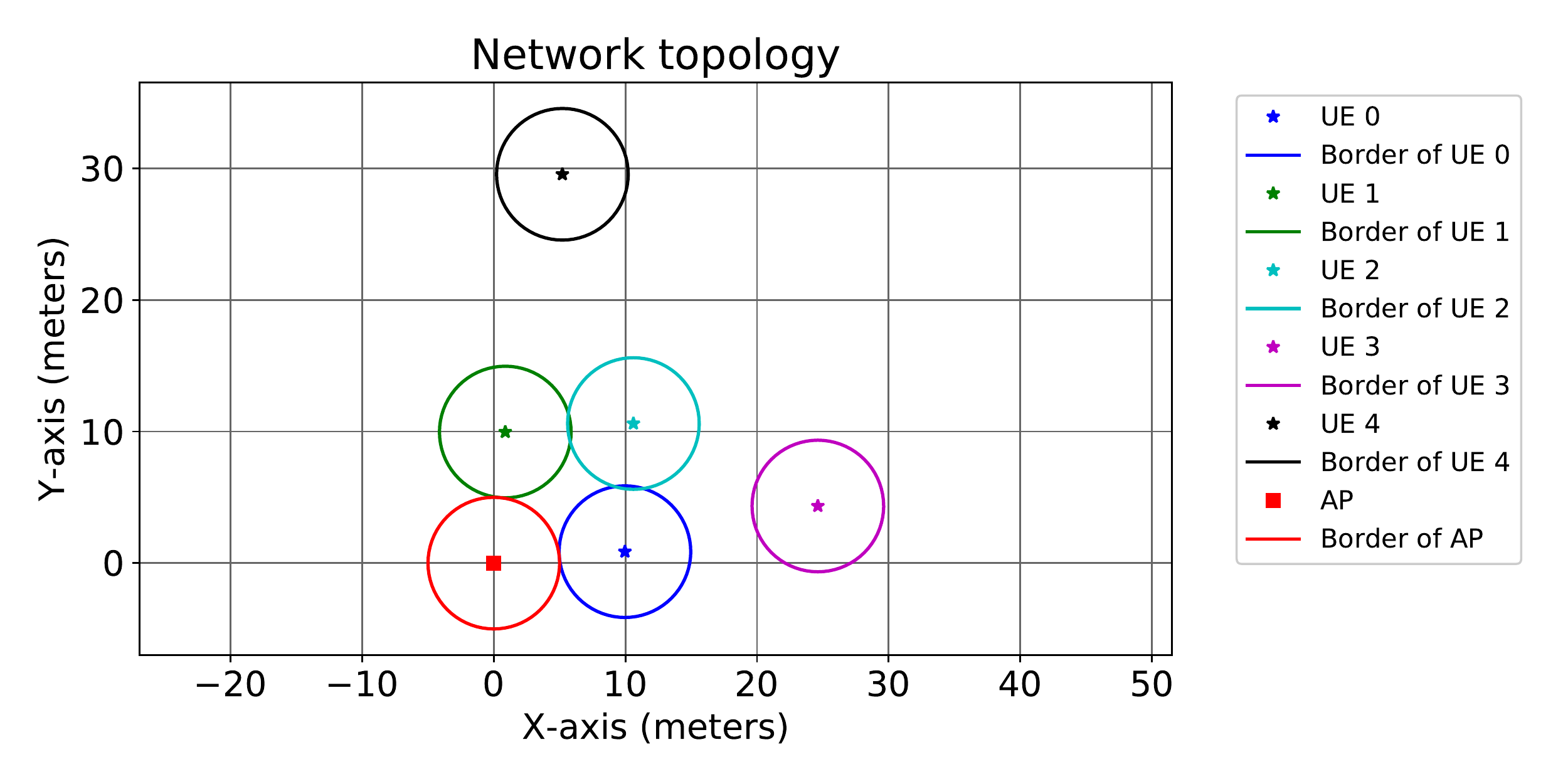}
		\captionsetup{type=figure,justification=centering}
		\caption{Network topology}
		\label{fig:network_topology1}
\end{figure}
\subsubsection{Description of scenario}
The network topology used for evaluation is shown in Fig.~\ref{fig:network_topology1}, where the initial position of the AP is set as the origin. The initial distances between the five UEs and the AP are $10, 10, 15, 25, 30$ meters, respectively, and the initial angles of the five UEs from the AP, with respect to the x-axis, are $5, 85, 45, 10, 80$ degrees, respectively. All UEs and the AP randomly move within their respective circular area of which the center is the initial position and the radius is 5 meters. The borders of the circular areas are shown for reference and indicated by the legend of Fig.~\ref{fig:network_topology1}.
The total downlink traffic is 1 Gbps and its allocation among UEs is 1/7, 3/7, 1/7, 1/7, 1/7. For the blockage model $p_{u,n}^\text{block}$ in Fig.~\ref{fig:blockage_model}, they are assigned valued with the following rules:
$p_{u,1}^{\text{block}}=0$ for $u\in[U]^+$;
$p_{u,n}^{\text{block}}=0.0026$ for $u\in\left\{1,2,4,5\right\}$ and $2\leq n\leq N^\text{block}$;
$p_{u,n}^{\text{block}}=0.1$ for $u\in\left\{3\right\}$ and $2\leq n\leq N^\text{block}$;
$p_{u,0}^{\text{block}}=1-\sum_{n=1}^{N^\text{block}}p_{u,n}^{\text{block}}$ for $u\in[U]^+$.
With the above assignments and the model in Fig. 3(b), 
the expected ratio of time that the main link of the UE is under blockage to the total time is $0.05, 0.05, 0.8, 0.05, 0.05$ for the five UEs, respectively. These value are defined as the \textit{probability of being under blockage} and denoted as $\tilde{p}_{u}^{\text{B}}$ for UE $u$. Note that these values are calculated by the numerical simulation for reference. Similarly, we use the model in Fig. 3(b) to characterize the blockage condition between any two UEs, with the probability of being under blockage for any D2D links being 0.05.
\subsubsection{Discussion on simulation parameter setup}
Given the complexity of the studied system, evaluating all potential scenarios would be demanding since there are many possibilities of network topology and data traffic patterns as multiple UEs are considered. The simulation setup described above has been inspired by the following intuition: (1) Network topology: different large-scale distances between AP and UEs would result in different channel quality among UEs, which is one of the key factors that motivate the joint optimization of user scheduling and link configuration. 
(2) User mobility: a restricted moving area per UE guarantees that their channel conditions will not change dramatically during the whole evaluation process. This is because the RL algorithms are generally effective when the statistics of the environment are stationary. Simulating a completely random system, meaning that the statistics are changing much faster than the convergence speed of the developed RL algorithms, is unfair in evaluating the RL algorithms. This is because once the environment has changed dramatically, the RL algorithms usually need to be reinitialized to recapture the system statistics. Therefore, we did not set all users randomly moving in the entire space, which is consistent with our system modeling. 
(3) Data traffic: the uneven packet arrival rates among UE aim at testing whether the proposed RL-based controllers could learn to use the UEs with low traffic demands as a relay for the other UEs with larger traffic demands. 
(4) Other parameters shown in Table~\ref{tab:sys_paras} are set with reasonable values according to literature and wireless standards. 
More simulation results on other scenarios could be found in our open-source code due to space limitations~\cite{YiZhang_DeepRL_mmWave_MANET_Codes}.

\subsection{Performance comparisons}\label{sub:sim_per_comp}
\subsubsection{Evolution of performance during whole monitored time}
\begin{figure*}[t!]
	\begin{subfigure}[t]{0.5\textwidth}
		\centering
		\includegraphics[width = 0.99\textwidth]{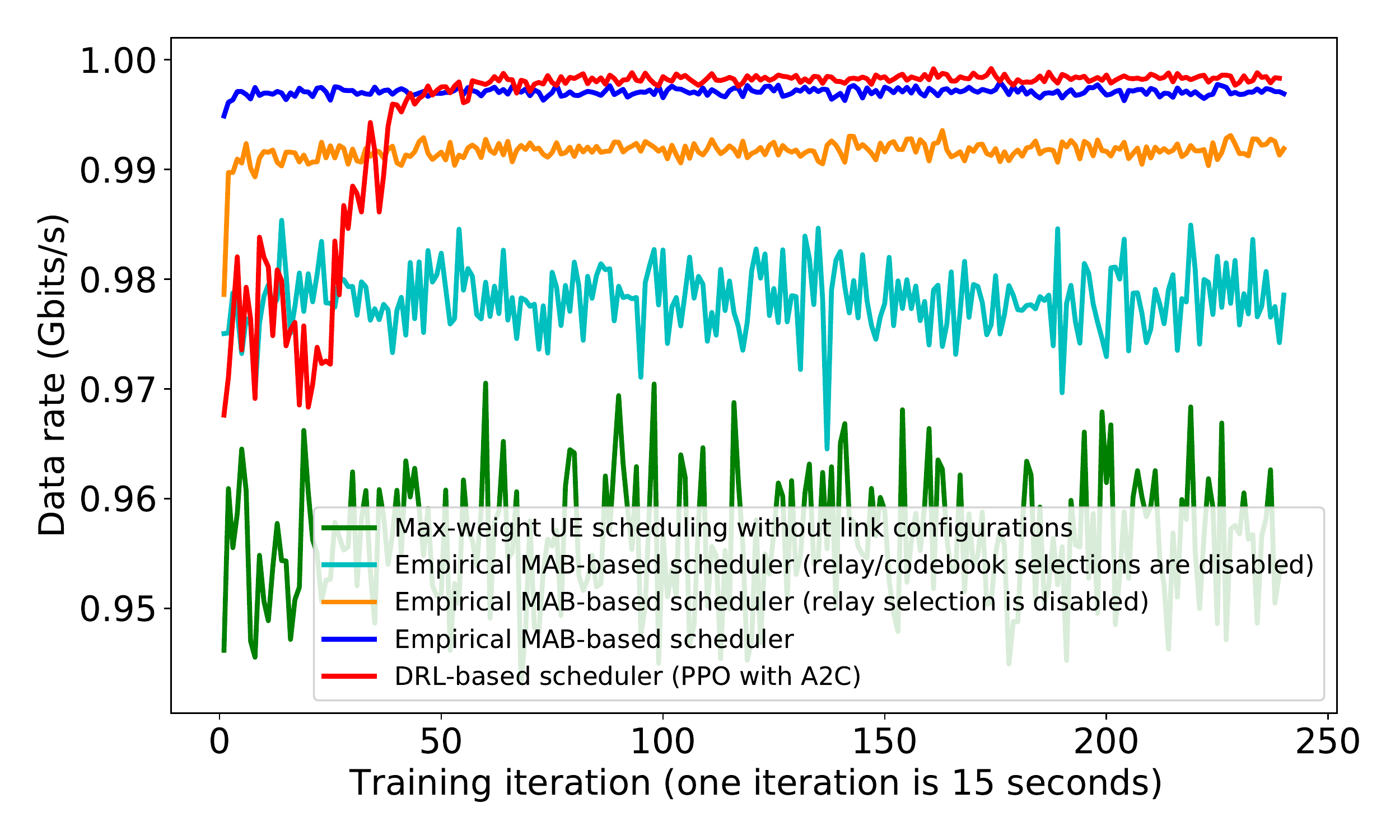}
		\captionsetup{type=figure,justification=centering}
		\caption{System overall data rate}
		\label{fig:Training_phase_evolution_rate_ckpt_3}
	\end{subfigure}
	\begin{subfigure}[t]{0.5\textwidth}
	 	\centering
		\includegraphics[width = 0.99\textwidth]{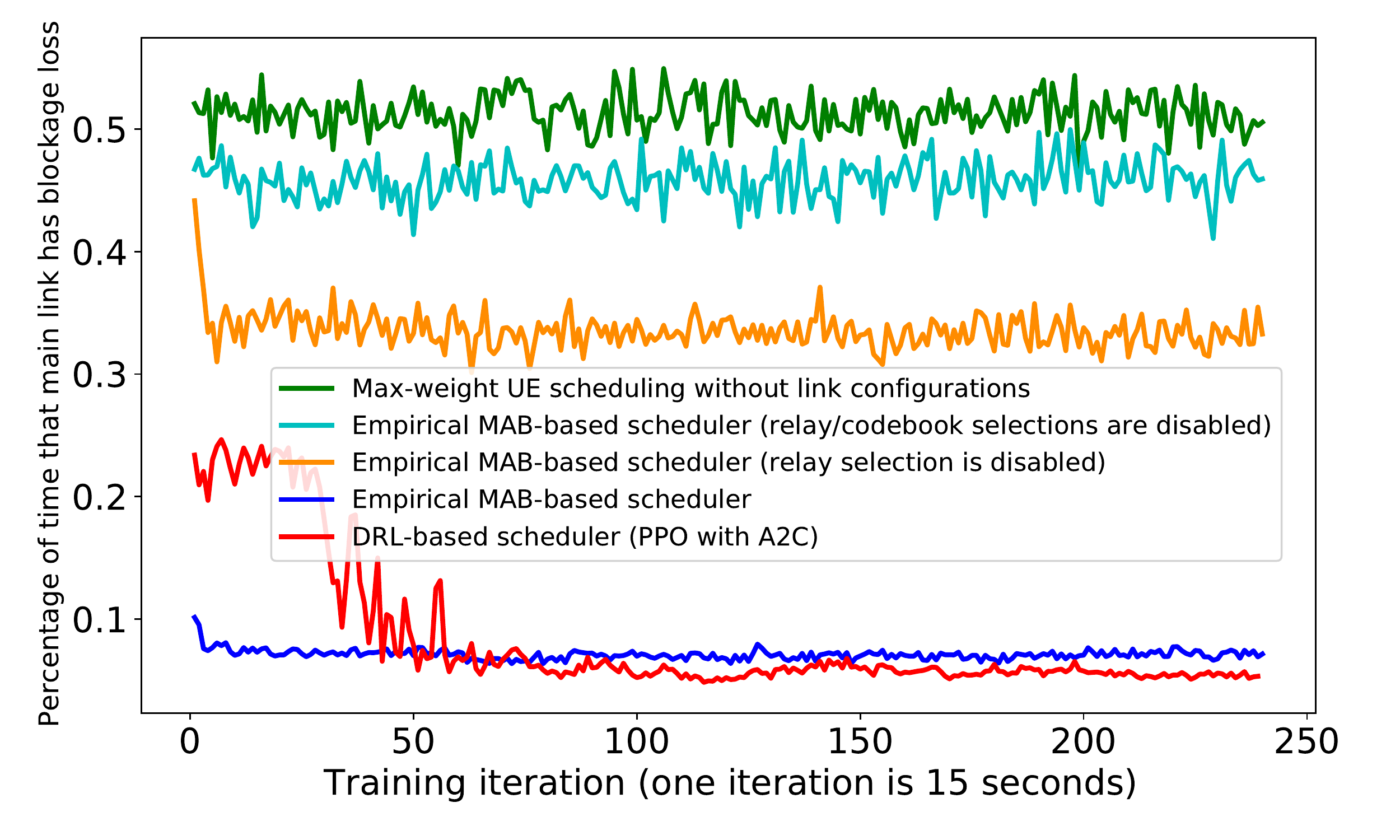}
		\captionsetup{type=figure,justification=centering}
		\caption{Percentage of time that main link is under blockage}
		\label{fig:Training_phase_evolution_ratio_blockage_ckpt_3}
	\end{subfigure}
	\begin{subfigure}[t]{0.5\textwidth}
		\centering
		\includegraphics[width = 0.99\textwidth]{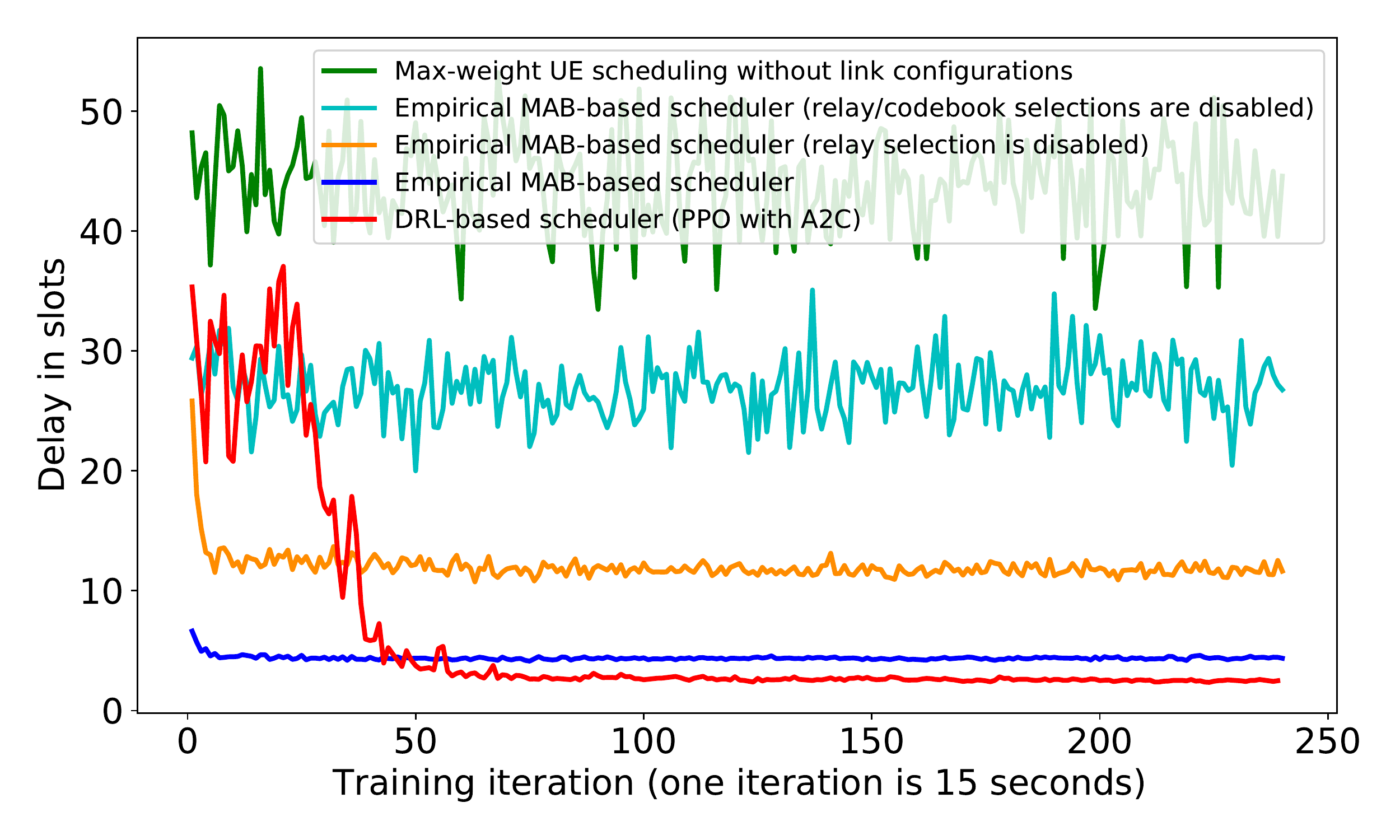}
		\captionsetup{type=figure,justification=centering}
		\caption{Average delay per packet}
		\label{fig:Training_phase_evolution_delay_ckpt_3}
	\end{subfigure}
	\begin{subfigure}[t]{0.5\textwidth}
		\centering
		\includegraphics[width = 0.99\textwidth]{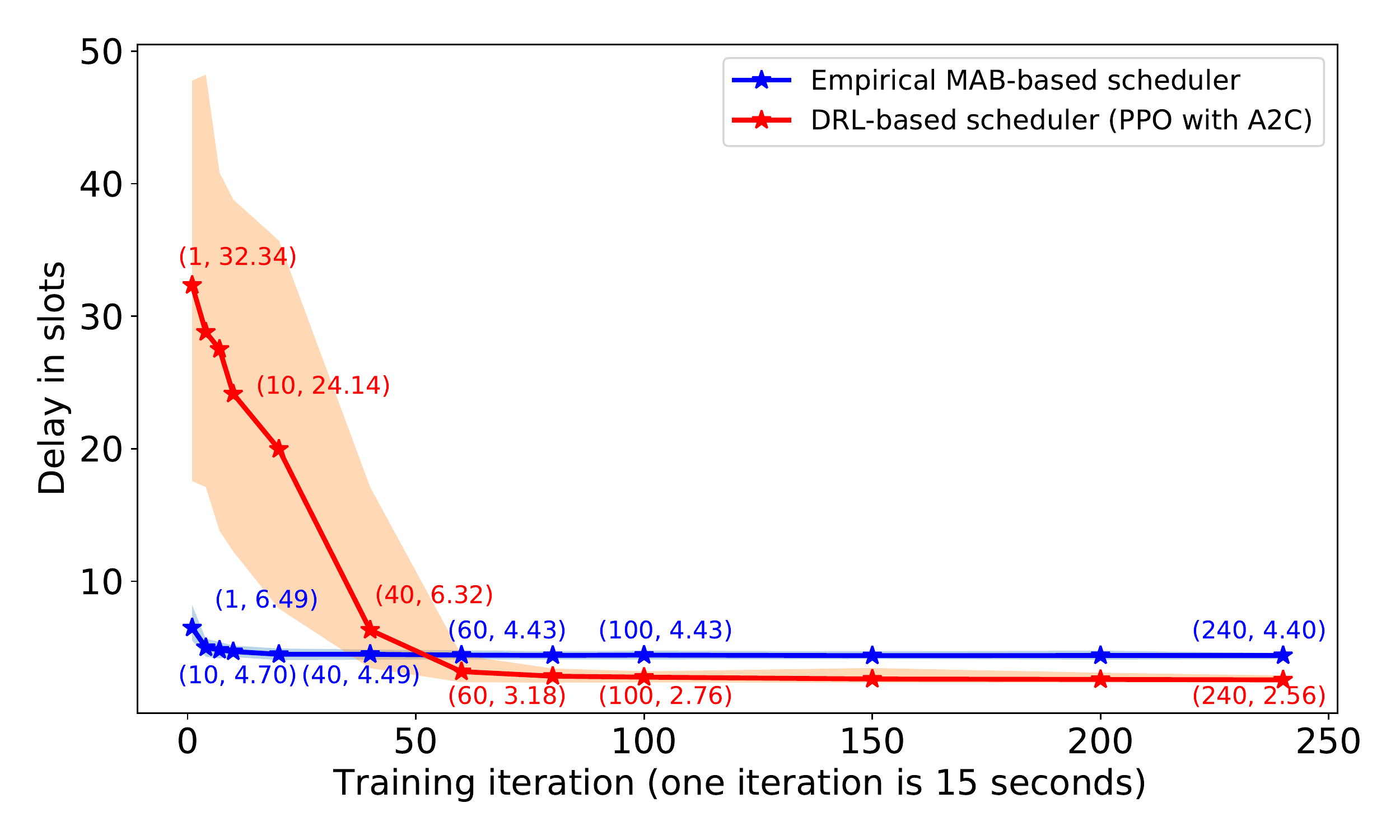}
		\captionsetup{type=figure,justification=centering}
		\caption{Performance robustness to stochastic training}
		\label{fig:Ave_Training_phase_evolution_delay_ckpt}
	\end{subfigure}
	\caption{Performance versus training iterations}
	\label{fig:Training_phase_performance_3}
\end{figure*}

In Fig.~\ref{fig:Training_phase_performance_3}, we show the evolution of the performance of the learned DRL-based controller and Empirical MAB-based controllers versus the training iteration. We recall that one iteration consists of 1500 time slots.
In particular, each data point in Fig.~\ref{fig:Training_phase_evolution_rate_ckpt_3} to Fig.~\ref{fig:Training_phase_evolution_delay_ckpt_3} is obtained by averaging the testing results of the learned controllers for 20 realizations and the each realization has 1500 time slots.
From Fig.~\ref{fig:Training_phase_performance_3}, several observations can be summarized as below:
(1) Fig.~\ref{fig:Training_phase_evolution_rate_ckpt_3} presents the evolution of the overall data rate, which is bounded by the arrival rate $\sum_{u=1}^U\lambda_u S^\text{pkg} =1$ Gbps. A stable policy should provide a data rate that equals the arrival rate, otherwise, the queue would explode. Fig.~\ref{fig:Training_phase_evolution_rate_ckpt_3} shows that the Empirical MAB-based controllers without relay selection all suffer from a data rate smaller than 1 Gbps, which yields exploding queue networks. This is because without using a relay, UE 3 is frequently under blockage and its direct link has very poor quality, which is not able to support the required data traffic.
(2) Fig.~\ref{fig:Training_phase_evolution_ratio_blockage_ckpt_3} shows that the proposed two controllers can learn good policies to select relays for the blocked UEs such that the average percentage of time that the UEs are suffering from the blockage loss in a main link is brought down to below 10\%. In particular, the DRL-based controller makes this percentage even below 5\%.
(3) Fig.~\ref{fig:Training_phase_evolution_delay_ckpt_3} provides the performance of the average delay per packet. We can see that both controllers provide acceptable delay performance that the average delay is within 5 time slots (50 ms). In particular, the DRL-based controller makes delay even below 30 ms.
(4) Note that Fig.~\ref{fig:Training_phase_evolution_rate_ckpt_3} to Fig.~\ref{fig:Training_phase_evolution_delay_ckpt_3} only monitored one training process. In Fig.~\ref{fig:Ave_Training_phase_evolution_delay_ckpt}, 
we show the robustness of the proposed algorithms to the stochastic training by averaging the results over 60 training processes. The area between max and min values of the 60 training processes is shaded in Fig.~\ref{fig:Ave_Training_phase_evolution_delay_ckpt}. We can observe that it overall took around 60 iterations (15 minutes) for the DRL-based controller to learn a good policy and its final performance is better than that of the Empirical MAB-based controller. The Empirical MAB-based controller only requires 10 iterations (2.5 minutes) to learn a good enough policy.

\subsubsection{Testing performance of the learned controllers}
\begin{figure*}[t!]
	\begin{subfigure}[t]{0.5\textwidth}
		\centering
		\includegraphics[width = 0.99\textwidth]{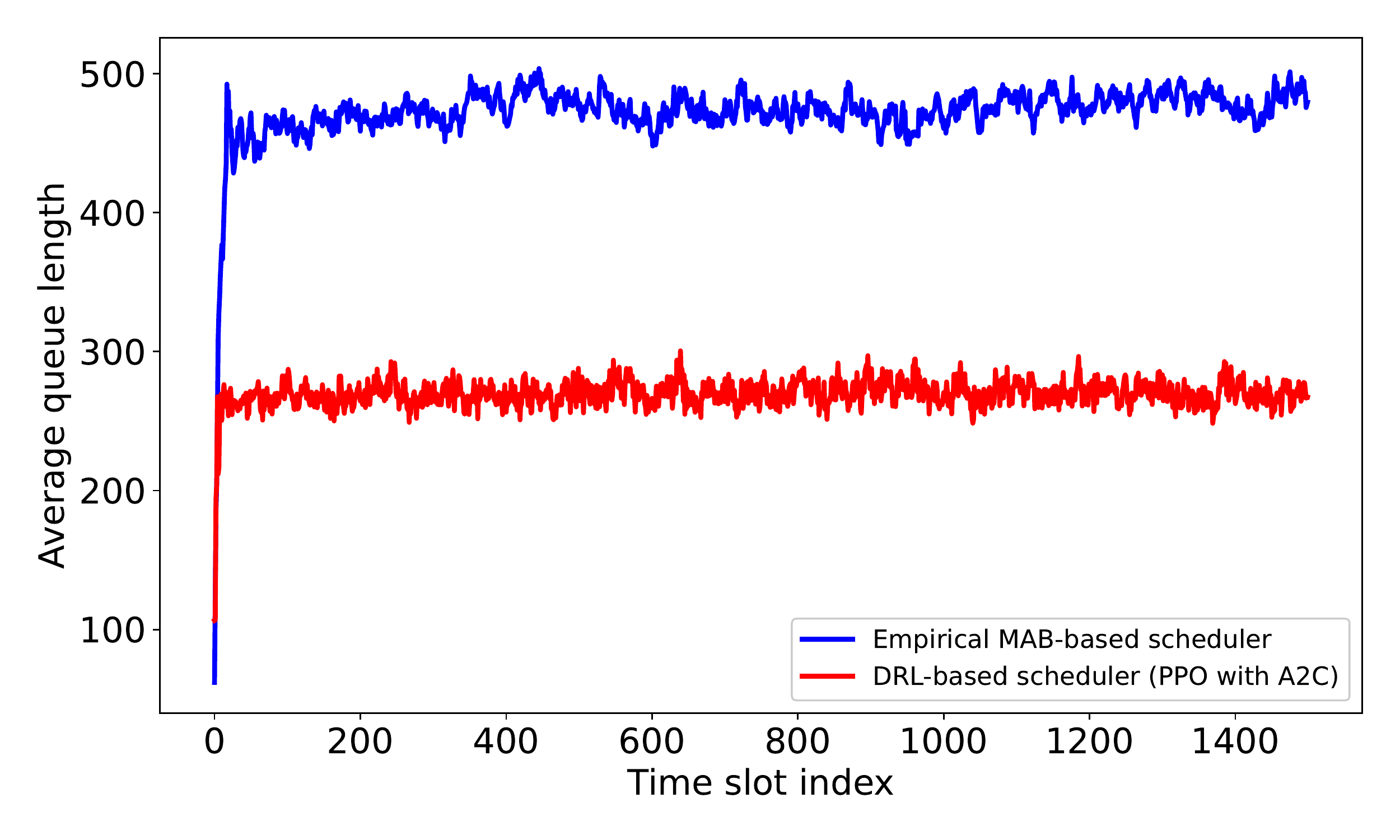}
		\captionsetup{type=figure,justification=centering}
		\caption{Evolution of average queue length per UE}
		\label{fig:Testing_phase_evolution_queue_length_3}
	\end{subfigure}
	\begin{subfigure}[t]{0.5\textwidth}
		\centering
		\includegraphics[width = 0.99\textwidth]{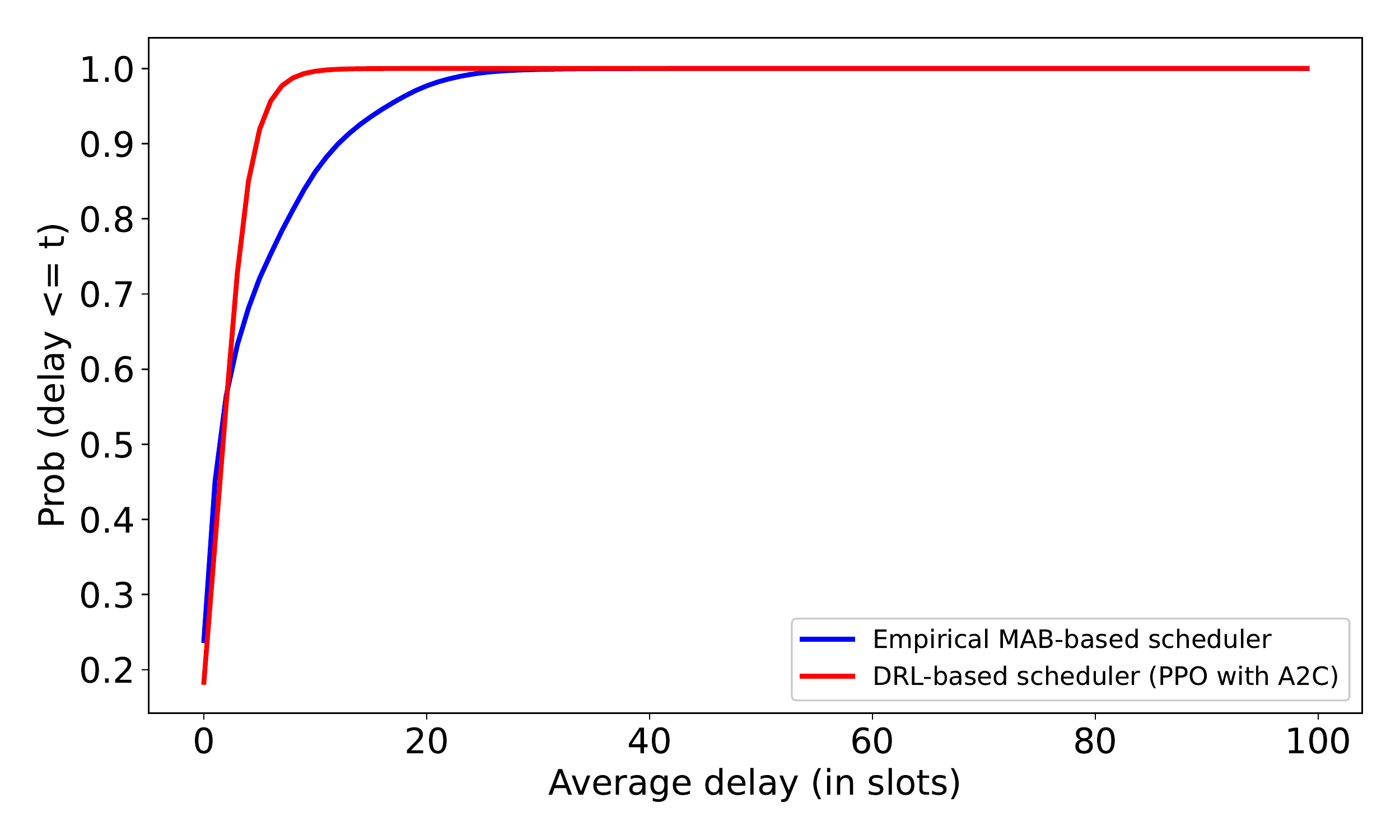}
		\captionsetup{type=figure,justification=centering}
		\caption{CDF of delay per packet}
		\label{fig:Testing_phase_CDF_delay_3}
	\end{subfigure}
	\caption{Testing performance of trained controllers}
	\label{fig:Testing_phase_3}
\end{figure*}

We now provide the testing performance of the proposed controllers (trained with 240 iterations) in Fig.~\ref{fig:Testing_phase_3}.
The result is an average of 200 realizations and each realization consists of 1500 time slots. The evolution of queue length is given in Fig.~\ref{fig:Testing_phase_evolution_queue_length_3} and the cumulative distribution function (CDF) of the delay per packet is shown in Fig.~\ref{fig:Testing_phase_CDF_delay_3}. It can be seen that both learned controllers well stabilize the queue. 
The average delay provided by the DL-based controller and the Empirical MAB-based controller is 25.6 ms and 44.0 ms, respectively.
Combined with the results given in Fig.~\ref{fig:Training_phase_performance_3}, we can conclude that the DRL-based controller provides better performance in terms of system delay but it requires higher sample complexity (roughly 6 times) as around 60 iterations are required for the DRL-based controller while only 10 iterations are enough for the MAB-based controller.
It is worth noting that the statistics of the system are the same for both the training and testing phase. If the statistics, most importantly the network topology of users, have changed dramatically, the agent has to be retrained. For example, the optimal codebook depends on UE mobility and its distance to the AP. If the distance changed dramatically, e.g. from being close to being far from the AP, the optimal codebook could be changed from being wide to being narrow. One potential solution is to retrain the system with a previously learned agent as an initial point.

\subsubsection{Complementary simulation results}
\begin{figure*}[t!]
	\centering
	\includegraphics[width = 0.98\textwidth]{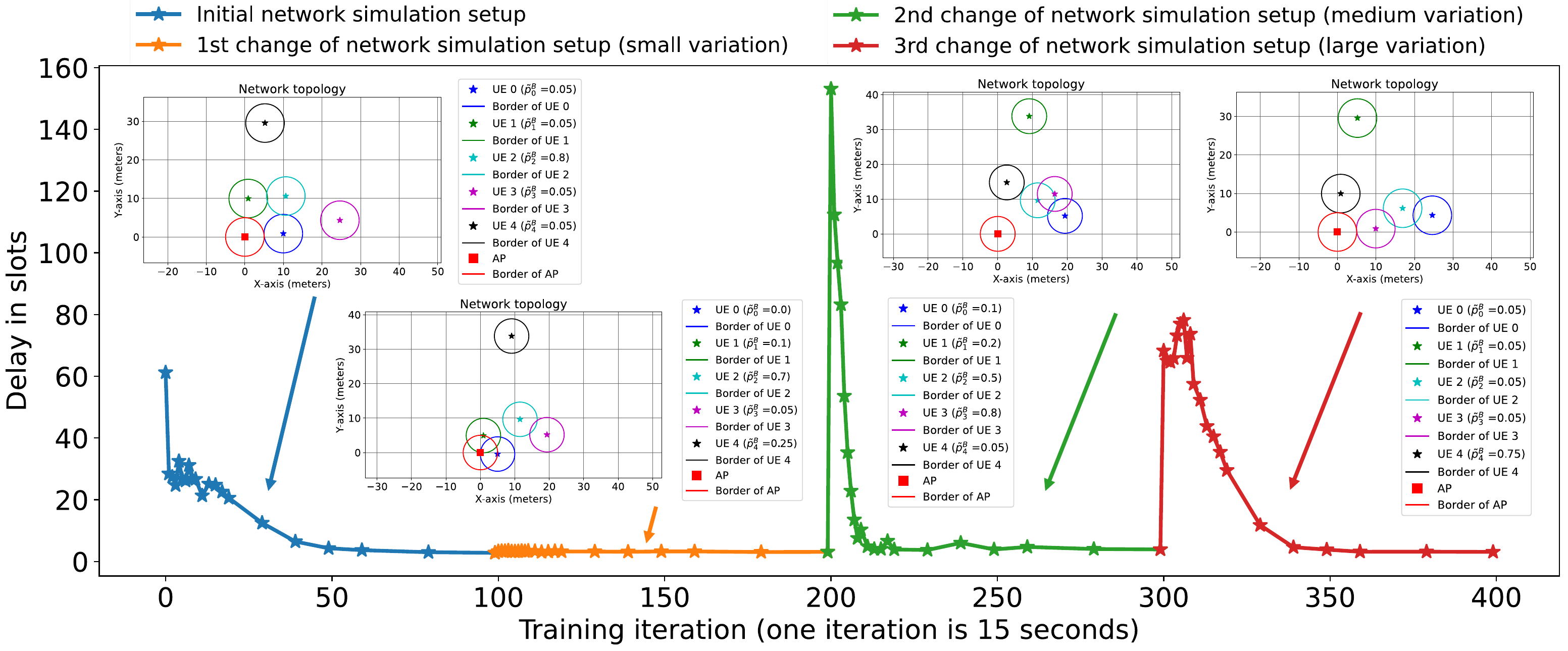}
	\captionsetup{type=figure,justification=centering}
	\caption{Average delay per packet versus training iterations (different network typologies)}
	\label{fig:changing_netw}
\end{figure*}
\begin{figure*}
	\begin{minipage}[t]{0.50\textwidth}
	\centering
	\includegraphics[width = 0.98\textwidth]{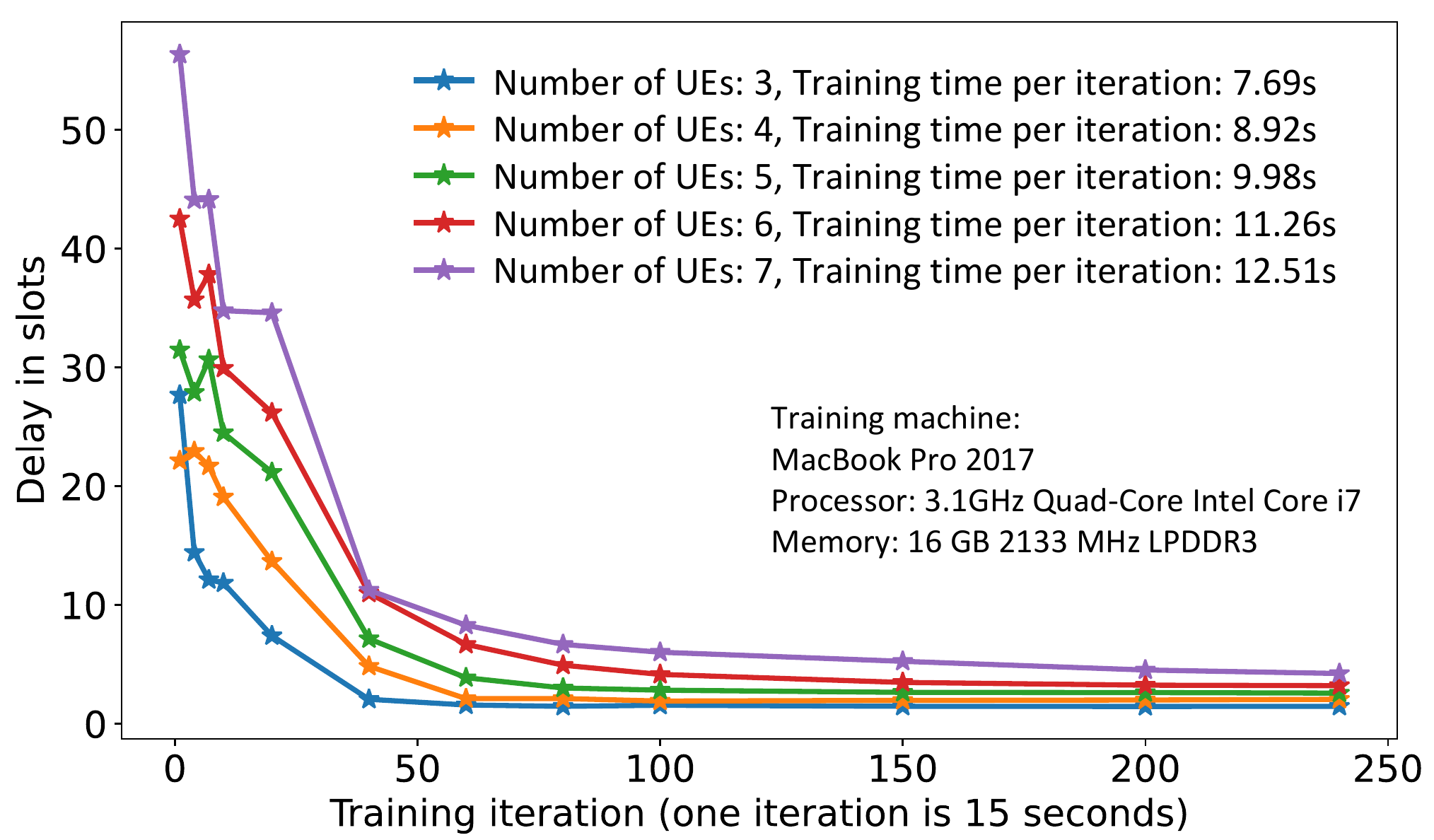}
	\captionsetup{type=figure,justification=centering}
	\caption{Average delay per packet versus training iterations (different number of UEs)}
	\label{fig:changing_ue}
	\end{minipage}
	\begin{minipage}[t]{0.50\textwidth}
	\centering
	\includegraphics[width = 0.98\textwidth]{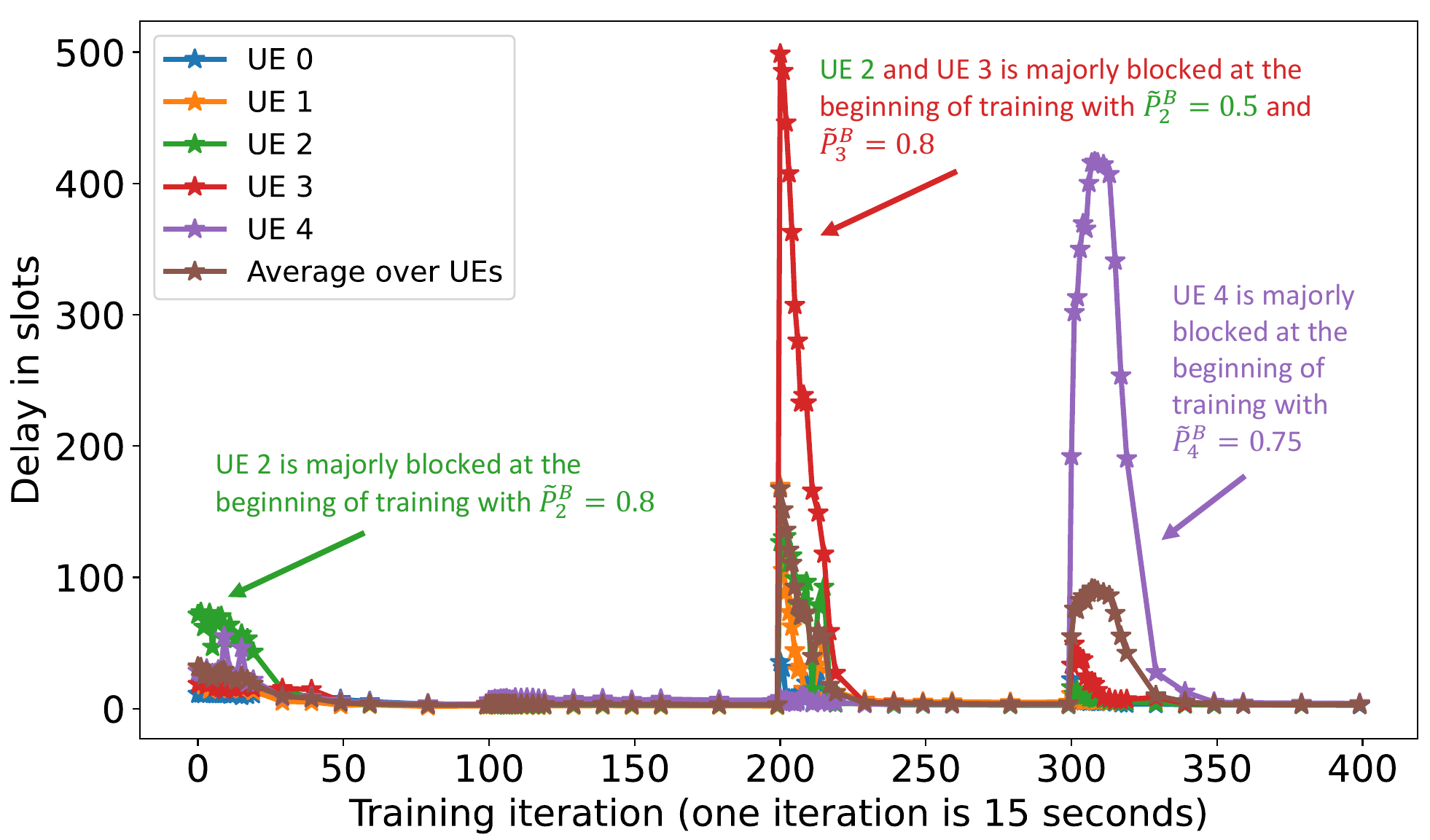}
	\captionsetup{type=figure,justification=centering}
	\caption{Average delay per packet versus training iterations (performance per UE)}
	\label{fig:show_each_ue}
	\end{minipage}
\end{figure*}
In this part, we provide several complementary simulation results. In particular, the results show in Fig.~\ref{fig:changing_netw}-Fig.~\ref{fig:changing_ue} are average over 10 training processes. For more detailed simulation setups, please see our open-source code~\cite{YiZhang_DeepRL_mmWave_MANET_Codes}.
In Fig.~\ref{fig:changing_netw}, we show the delay performance provided by the DRL-based controller over 400 training iterations, where the network simulation setup is changed every 100 iterations, hence being able to evaluate the ``transfer learning" capability of the proposed DRL-based controller for the non-stationary environment. In particular, the 1st network change is only a small variation of the UE positions, which is referred to as a ``small variation" in the legend. The 2nd network change consists of a considerable variation of the UE positions and a small variation of the UE blockage parameters, which is referred to as a ``medium variation" in the legend. The 3rd network change consists of a considerable variation of the UE positions and a considerable variation of the UE blockages parameters, which is referred to as a ``large variation". The network typologies and the probability of being under blockage for each UE, i.e. $\left\{\tilde{p}_{u}^{\text{B}}\right\}$, are also provided in the figure. As shown in Fig~\ref{fig:changing_netw}, a small variation of the environment setup has little impact on the learned DRL-based controller, while a larger variation might require 10 iterations (for the 2nd change) or even 40 iterations (for the 3rd change) to relearn the environment. This experiment confirms that the proposed DRL-based algorithm did learn a certain amount of meaningful features from the relationship of state-action-reward, hence being able to potentially deal with the non-stationary environments.
In Fig.~\ref{fig:show_each_ue}, we further break down the delay performance shown in Fig.~\ref{fig:changing_netw} into each UE perspective. As we can see, the large delay at the beginning of the training process is majorly due to certain UE(s) that are under blockage. Once a good scheduling policy is learned, the delay is significantly reduced and the queue becomes stable.
In Fig.~\ref{fig:changing_ue}, we further show the performance versus training iteration and the computational time per iteration for different numbers of UEs. It is worth pointing out that the absolute value of the training time per iteration depends on the running machine.

\section{Conclusions}\label{sec:discussion}
In this work, we studied the joint user scheduling and link configuration in a multi-user mmWave network.
We modeled this complex scheduling/designing problem as a dynamic decision-making process and proposed two RL-based solutions. Our evaluation confirmed their viability and effectiveness and also showed that the DRL-based solution provides better system performance while the MAB-based solution results in a faster training process.
One potential future direction is to exploit multiple mmWave array groups and OFDMA to further reduce the link latency by simultaneously communicating with multiple users.

\bibliographystyle{IEEEtran}
\bibliography{ref-DRLSch-all-abrv}

\begin{thebibliography}{10}
\providecommand{\url}[1]{#1}
\csname url@samestyle\endcsname
\providecommand{\newblock}{\relax}
\providecommand{\bibinfo}[2]{#2}
\providecommand{\BIBentrySTDinterwordspacing}{\spaceskip=0pt\relax}
\providecommand{\BIBentryALTinterwordstretchfactor}{4}
\providecommand{\BIBentryALTinterwordspacing}{\spaceskip=\fontdimen2\font plus
\BIBentryALTinterwordstretchfactor\fontdimen3\font minus
  \fontdimen4\font\relax}
\providecommand{\BIBforeignlanguage}[2]{{%
\expandafter\ifx\csname l@#1\endcsname\relax
\typeout{** WARNING: IEEEtran.bst: No hyphenation pattern has been}%
\typeout{** loaded for the language `#1'. Using the pattern for}%
\typeout{** the default language instead.}%
\else
\language=\csname l@#1\endcsname
\fi
#2}}
\providecommand{\BIBdecl}{\relax}
\BIBdecl

\bibitem{Ted_mmWave_will_work_2013Access}
T.~S. Rappaport, S.~Sun, R.~Mayzus, H.~Zhao, Y.~Azar, K.~Wang, G.~N. Wong,
  J.~K. Schulz, M.~Samimi, and F.~Gutierrez, ``Millimeter wave mobile
  communications for 5g cellular: It will work!'' \emph{IEEE Access}, vol.~1,
  pp. 335--349, 2013.

\bibitem{Han_LargeScaleMmWaveComMag14}
S.~Han, C.~lin I, Z.~Xu, and C.~Rowell, ``Large-scale antenna systems with
  hybrid analog and digital beamforming for millimeter wave 5{G},''
  \emph{{IEEE} Commun. Mag.}, vol.~53, no.~1, pp. 186--194, 2015.

\bibitem{Multinomial_TS_BeamTracking_Aykin_Infocom20}
I.~Aykin, B.~Akgun, M.~Feng, and M.~Krunz, ``{MAMBA}: A multi-armed bandit
  framework for beam tracking in millimeter-wave systems,'' in \emph{Proc. of
  2020 IEEE Int. Conf. on Comput. Commun.}, Shanghai, China, 2020, pp.
  1469--1478.

\bibitem{Relay_selection_scheduling_MASS2017}
Q.~Hu and D.~M. Blough, ``Relay selection and scheduling for millimeter wave
  backhaul in urban environments,'' in \emph{Proc. of 2017 IEEE 14th Int. Conf.
  on Mobile Ad Hoc and Sensor Systems}, Orlando, FL, USA, 2017, pp. 206--214.

\bibitem{Outage_duration_TVT2021}
C.~Tunc, M.~F. \"Ozkoç, F.~Fund, and S.~S. Panwar, ``The blind side: Latency
  challenges in millimeter wave networks for connected vehicle applications,''
  \emph{{IEEE} Trans. Veh. Technol.}, vol.~70, no.~1, pp. 529--542, Jan. 2021.

\bibitem{Relay_mmWave_ComMag2017}
J.~Deng, O.~Tirkkonen, R.~Freij-Hollanti, T.~Chen, and N.~Nikaein, ``Resource
  allocation and interference management for opportunistic relaying in
  integrated mm{W}ave/sub-6 {GHz} 5{G} networks,'' \emph{{IEEE} Commun. Mag.},
  vol.~55, no.~6, pp. 94--101, Jun. 2017.

\bibitem{Relay_probing_MmWave_Wei_TVT16}
N.~Wei, X.~Lin, and Z.~Zhang, ``Optimal relay probing in millimeter-wave
  cellular systems with device-to-device relaying,'' \emph{{IEEE} Trans. Veh.
  Technol.}, vol.~65, no.~12, pp. 10\,218--10\,222, Dec. 2016.

\bibitem{Blockage_scheduling_WPAN_Niu_TVT15}
Y.~Niu, Y.~Li, D.~Jin, L.~Su, and D.~Wu, ``Blockage robust and efficient
  scheduling for directional mm{W}ave {WPAN}s,'' \emph{{IEEE} Trans. Veh.
  Technol.}, vol.~64, no.~2, pp. 728--742, Feb. 2015.

\bibitem{802.11ad_standard:2016}
``{IEEE} {S}tandard for information technology--telecommunications and
  information exchange between systems local and metropolitan area
  networks--specific requirements - part 11: {W}ireless {LAN} medium access
  control ({MAC}) and physical layer ({PHY}) specifications,'' \emph{{IEEE} Std
  802.11-2016 (Revision of IEEE Std 802.11-2012)}, pp. 1--3534, Dec. 2016.

\bibitem{11ay_standard_2019}
``{IEEE} {Draft} {Standard} for information technology--telecommunications and
  information exchange between systems local and metropolitan area
  networks--specific requirements part 11: Wireless {LAN} medium access control
  ({MAC}) and physical layer ({PHY}) specifications--amendment: Enhanced
  throughput for operation in license-exempt bands above 45 {GHz},''
  \emph{{IEEE} P802.11ay/D4.0, June 2019}, pp. 1--791, Jul. 2019.

\bibitem{5G_NR_standard_V16:2019}
``System architecture for the 5{G} system,'' \emph{document {TS} 23.501
  V16.1.0, {3GPP}, Jun. 2019}, pp. 1--219, Jun. 2019.

\bibitem{MUTE_Yasaman_Mobicom18}
Y.~Ghasempour, M.~K. Haider, C.~Cordeiro, D.~Koutsonikolas, and E.~Knightly,
  ``Multi-stream beam-training for mm{W}ave {MIMO} networks,'' in \emph{Proc.
  of the 24th Annu. Int. Conf. on Mobile Comput. and Netw.}, New Delhi, India,
  2018, pp. 225--239.

\bibitem{ThoHea:Ergodic-Rate-AdHoc:18}
A.~Thornburg and R.~W. Heath, ``Ergodic rate of millimeter wave ad hoc
  networks,'' \emph{{IEEE} Trans. on Wireless Commun.}, vol.~17, no.~2, pp.
  914--926, Feb. 2018.

\bibitem{Lee_BA_UE_scheduling_with_Mobility_Wiopt19}
J.~Lee and E.~Ekici, ``Beam alignment and user scheduling in mm{W}ave networks
  under mobility,'' in \emph{Proc. of 2019 Int. Symp. on Modeling and
  Optimization in Mobile, Ad Hoc, and Wireless Networks}, Avignon, France, Jun.
  2019, pp. 1--8.

\bibitem{TSCB_YiZhang_Mobihoc_2021}
Y.~Zhang, S.~Basu, S.~Shakkottai, and R.~W. {Heath Jr.}, ``{MmWave} codebook
  selection in rapidly-varying channels via multinomial {T}hompson sampling,''
  in \emph{Proc. of the 22nd ACM Int. Symp. on Mobile Ad Hoc Netw. and
  Comput.}, Shanghai, China, 2021, pp. 151--160.

\bibitem{V2X_mmWave_measurements_Mobicom20}
S.~Wang, J.~Huang, and X.~Zhang, ``Demystifying millimeter-wave {V2X}: Towards
  robust and efficient directional connectivity under high mobility,'' in
  \emph{Proc. of the 26th Annu. Int. Conf. on Mobile Comput. and Netw.},
  London, United Kingdom, 2020.

\bibitem{RL_Intro_Sutton_book18}
R.~S. Sutton and A.~G. Barto, \emph{Reinforcement Learning: An
  Introduction}.\hskip 1em plus 0.5em minus 0.4em\relax Cambridge, MA, USA: The
  MIT Press, 2018.

\bibitem{DRL_wireless_overview_VTMag19}
Z.~Xiong, Y.~Zhang, D.~Niyato, R.~Deng, P.~Wang, and L.-C. Wang, ``Deep
  reinforcement learning for mobile 5{G} and beyond: Fundamentals,
  applications, and challenges,'' \emph{IEEE Veh. Technol. Mag.}, vol.~14,
  no.~2, pp. 44--52, Jun. 2019.

\bibitem{Luong_DRL_wireless_ComSurvey_19}
N.~C. Luong, D.~T. Hoang, S.~Gong, D.~Niyato, P.~Wang, Y.-C. Liang, and D.~I.
  Kim, ``Applications of deep reinforcement learning in communications and
  networking: A survey,'' \emph{{IEEE} Commun. Surveys Tuts.}, vol.~21, no.~4,
  pp. 3133--3174, May 2019.

\bibitem{PPO_Schulman_17}
J.~Schulman, F.~Wolski, P.~Dhariwal, A.~Radford, and O.~Klimov, ``Proximal
  policy optimization algorithms,'' \emph{arXiv:1707.06347}, 2017.

\bibitem{A3C_Mnih_ICML16}
V.~Mnih, A.~P. Badia, M.~Mirza, A.~Graves, T.~Lillicrap, T.~Harley, D.~Silver,
  and K.~Kavukcuoglu, ``Asynchronous methods for deep reinforcement learning,''
  in \emph{Proc. of The 33rd Int. Conf. on Machine Learning}, vol.~48, New York
  City, New York, USA, 20--22 Jun 2016, pp. 1928--1937.

\bibitem{MaxWeight_TIT_93}
L.~Tassiula and A.~Ephremides, ``Dynamic server allocation to parallel queues
  with randomly varying connectivity,'' \emph{{IEEE} Trans. Inf. Theory},
  vol.~39, no.~2, pp. 466--478, 1993.

\bibitem{Delay_analysis_maxweight}
M.~J. Neely, ``Delay analysis for max weight opportunistic scheduling in
  wireless systems,'' \emph{{IEEE} Trans. Automat. Contr.}, vol.~54, no.~9, pp.
  2137--2150, Sept. 2009.

\bibitem{DDPG_Sch_Min_Delay_LTE_PIMRC_2020}
N.~Sharma, S.~Zhang, S.~R. Somayajula~Venkata, F.~Malandra, N.~Mastronarde, and
  J.~Chakareski, ``Deep reinforcement learning for delay-sensitive {LTE}
  downlink scheduling,'' in \emph{2020 IEEE 31st Annu. Int. Symp. on Personal,
  Indoor and Mobile Radio Commun.}, London, UK, 2020, pp. 1--6.

\bibitem{DQN_Mnih_Nature15}
V.~Mnih, K.~Kavukcuoglu, D.~Silver, A.~A. Rusu, J.~Veness, M.~G. Bellemare, and
  et~al, ``Human-level control through deep reinforcement learning,''
  \emph{Nature}, vol. 518, no. 7540, pp. 529--533, Feb. 2015.

\bibitem{DDPG_Lillicrap_ICLR16}
T.~P. Lillicrap, J.~J. Hunt, A.~Pritzel, N.~Heess, T.~Erez, Y.~Tassa,
  D.~Silver, and D.~Wierstra, ``Continuous control with deep reinforcement
  learning,'' in \emph{Proc. of the 4th Int. Conf. on Learning
  Representations}, San Juan, Puerto Rico, May 2016.

\bibitem{LSTM_DQN_Min_Delay_Globecom_2018}
M.~Elsayed and M.~Erol-Kantarci, ``Deep reinforcement learning for reducing
  latency in mission critical services,'' in \emph{2018 IEEE Global Commun.
  Conf.}, Abu Dhabi, United Arab Emirates, 2018, pp. 1--6.

\bibitem{DQN_RNN_Min_Delay_Globecom_2020}
T.~Zhang, S.~Shen, S.~Mao, and G.-K. Chang, ``Delay-aware cellular traffic
  scheduling with deep reinforcement learning,'' in \emph{Proc. of 2020 IEEE
  Global Commun. Conf.}, Taipei, Taiwan, 2020, pp. 1--6.

\bibitem{Intel_DRL_OFDMA_RA_Globecom19}
R.~Balakrishnan, K.~Sankhe, V.~S. Somayazulu, R.~Vannithamby, and J.~Sydir,
  ``Deep reinforcement learning based traffic- and channel-aware {OFDMA}
  resource allocation,'' in \emph{Proc. of 2019 IEEE Global Commun. Conf.},
  Waikoloa, HI, USA, Dec. 2019, pp. 1--6.

\bibitem{Gupta_DRL_link_schedule_TWC20}
M.~Gupta, A.~Rao, E.~Visotsky, A.~Ghosh, and J.~G. Andrews, ``Learning link
  schedules in self-backhauled millimeter wave cellular networks,''
  \emph{{IEEE} Trans. on Wireless Commun.}, vol.~19, no.~12, pp. 8024--8038,
  Dec. 2020.

\bibitem{Relay_probing_overview_mmWave_Access2020}
E.~M. Mohamed, B.~M. Elhalawany, H.~S. Khallaf, M.~Zareei, A.~Zeb, and M.~A.
  Abdelghany, ``Relay probing for millimeter wave multi-hop d2d networks,''
  \emph{IEEE Access}, vol.~8, pp. 30\,560--30\,574, 2020.

\bibitem{Predictive_Prelay_selection_ACCESS2019}
A.~Dimas, D.~S. Kalogerias, and A.~P. Petropulu, ``Cooperative beamforming with
  predictive relay selection for urban mm{W}ave communications,'' \emph{IEEE
  Access}, vol.~7, pp. 157\,057--157\,071, 2019.

\bibitem{Zhang_DRL_relay_power_selection_WCL20}
H.~Zhang, S.~Chong, X.~Zhang, and N.~Lin, ``A deep reinforcement learning based
  {D2D} relay selection and power level allocation in mm{W}ave vehicular
  networks,'' \emph{{IEEE} Wireless Commun. Lett.}, vol.~9, no.~3, pp.
  416--419, Mar. 2020.

\bibitem{SCAROS_epsilon_greedy_backhauling_routing_JSAC_2019}
A.~Ortiz, A.~Asadi, G.~H. Sim, D.~Steinmetzer, and M.~Hollick, ``{SCAROS}: A
  scalable and robust self-backhauling solution for highly dynamic
  millimeter-wave networks,'' \emph{{IEEE} J. Sel. Areas Commun.}, vol.~37,
  no.~12, pp. 2685--2698, Dec. 2019.

\bibitem{MU_BW_opt_Hossein_ICC_2015}
H.~Shokri-Ghadikolaei, L.~Gkatzikis, and C.~Fischione, ``Beam-searching and
  transmission scheduling in millimeter wave communications,'' in \emph{Proc.
  of 2015 IEEE Int. Conf. on Commun.}, London, UK, 2015, pp. 1292--1297.

\bibitem{Liu_JSAC_Wiopt_2019}
J.~Liu and E.~S. Bentley, ``Hybrid-beamforming-based millimeter-wave cellular
  network optimization,'' \emph{{IEEE} J. Sel. Areas Commun.}, vol.~37, no.~12,
  pp. 2799--2813, Dec. 2019.

\bibitem{DL_Learning_beam_pairs_Wang_Access_2019}
Y.~Wang, A.~Klautau, M.~Ribero, A.~C.~K. Soong, and R.~W. Heath, ``Mmwave
  vehicular beam selection with situational awareness using machine learning,''
  \emph{IEEE Access}, vol.~7, pp. 87\,479--87\,493, 2019.

\bibitem{FilSciDevCap_GeoDataBase_TMC}
I.~Filippini, V.~Sciancalepore, F.~Devoti, and A.~Capone, ``Fast cell discovery
  in mm-wave 5g networks with context information,'' \emph{{IEEE} Trans. Mobile
  Comput.}, vol.~17, no.~7, pp. 1538--1552, Jul. 2018.

\bibitem{Site_codebook_learning_Wang_Heath_TWC21}
Y.~Wang, N.~J. Myers, N.~Gonz\'{a}lez-Prelcic, and R.~W. Heath, ``Site-specific
  online compressive beam codebook learning in mmwave vehicular
  communication,'' \emph{{IEEE} Trans. on Wireless Commun.}, vol.~20, no.~5,
  pp. 3122--3136, Jan. 2021.

\bibitem{Jeong_OnlineTracking_CBSelection_Infocom20}
J.~Jeong, S.~H. Lim, Y.~Song, and S.-W. Jeon, ``Online learning for joint beam
  tracking and pattern optimization in massive mimo systems,'' in \emph{Proc.
  of 2020 IEEE Int. Conf. on Comput. Commun.}, Toronto, ON, Canada, 2020, pp.
  764--773.

\bibitem{alrabeiah_NN_beam_codebook_2020}
M.~Alrabeiah, Y.~Zhang, and A.~Alkhateeb, ``Neural networks based beam
  codebooks: Learning mmwave massive {MIMO} beams that adapt to deployment and
  hardware,'' \emph{arXiv: 2006.14501}, 2020.

\bibitem{Gao_DRL_bw_powerOptimization_CL20}
J.~Gao, C.~Zhong, X.~Chen, H.~Lin, and Z.~Zhang, ``Deep reinforcement learning
  for joint beamwidth and power optimization in mm{W}ave systems,''
  \emph{{IEEE} Commun. Lett.}, vol.~24, no.~10, pp. 2201--2205, Oct. 2020.

\bibitem{Palacios_Tracking_mmWaveChannel_Infocom17}
J.~Palacios, D.~De~Donno, and J.~Widmer, ``Tracking mm-{W}ave channel dynamics:
  Fast beam training strategies under mobility,'' in \emph{Proc. of 2017 IEEE
  Conf. on Comput. Commun.}, Atlanta, GA, USA, 2017, pp. 1--9.

\bibitem{WidebandChannelTracking_Nuria_TWC2021}
N.~Gonz\'alez-Prelcic, H.~Xie, J.~Palacios, and T.~Shimizu, ``Wideband channel
  tracking and hybrid precoding for mm{W}ave {MIMO} systems,'' \emph{{IEEE}
  Trans. on Wireless Commun.}, vol.~20, no.~4, pp. 2161--2174, Apr. 2021.

\bibitem{Shahram_Robust_BeamTracking_WiOPT19}
S.~Shahsavari, M.~A. Amir~Khojastepour, and E.~Erkip, ``Robust beam tracking
  and data communication in millimeter wave mobile networks,'' in \emph{Proc.
  of 2019 Int. Symp. on Modeling and Optimization in Mobile, Ad Hoc, and
  Wireless Networks}, Avignon, France, 2019, pp. 1--8.

\bibitem{CB_Design_Xiao_TWC_2017}
Z.~Xiao, P.~Xia, and X.-G. Xia, ``Codebook design for millimeter-wave channel
  estimation with hybrid precoding structure,'' \emph{{IEEE} Trans. on Wireless
  Commun.}, vol.~16, no.~1, pp. 141--153, Jan. 2017.

\bibitem{SANBA_YiZhang_Mobihoc_2019}
Y.~Zhang, K.~Patel, S.~Shakkottai, and R.~W. Heath~Jr.,
  ``Side-information-aided noncoherent beam alignment design for millimeter
  wave systems,'' in \emph{Proc. of the 20th ACM Int. Symp. on Mobile Ad Hoc
  Netw. and Comput.}, Catania, Italy, 2019, pp. 341--350.

\bibitem{Hse_mobility_model_manet_infocom07}
W.~Hsu, T.~Spyropoulos, K.~Psounis, and A.~Helmy, ``Modeling time-variant user
  mobility in wireless mobile networks,'' in \emph{Proc. of 2007 IEEE Int.
  Conf. on Comput. Commun.}, Anchorage, AK, USA, 2007, pp. 758--766.

\bibitem{Wang_blockage_wiopt17}
Y.~Wang and G.~de~Veciana, ``Temporal dynamics of mobile blocking in millimeter
  wave based wearable networks,'' in \emph{Proc. of 2017 Int. Symp. on Modeling
  and Optimization in Mobile, Ad Hoc, and Wireless Networks}, Paris, France,
  2017, pp. 1--8.

\bibitem{Schulman_TRPO_ICML15}
J.~Schulman, S.~Levine, P.~Abbeel, M.~Jordan, and P.~Moritz, ``Trust region
  policy optimization,'' in \emph{Proc. of the 32nd Int. Conf. on Machine
  Learning}, vol.~37, Lille, France, Jul. 2015, pp. 1889--1897.

\bibitem{GAE_Schulman_ICLR16}
J.~Schulman, P.~Moritz, S.~Levine, M.~I. Jordan, and P.~Abbeel,
  ``High-dimensional continuous control using generalized advantage
  estimation,'' in \emph{Proc. of the 4th Int. Conf. on Learning
  Representations}, San Juan, Puerto Rico, May 2016.

\bibitem{ZhuPOMDPs}
\BIBentryALTinterwordspacing
P.~Zhu, X.~Li, P.~Poupart, and G.~Miao, ``On improving deep reinforcement
  learning for {POMDP}s,'' 2017. [Online]. Available:
  \url{https://arxiv.org/abs/1704.07978}
\BIBentrySTDinterwordspacing

\bibitem{Empirical_TS_Chapelle_NIPS2011}
O.~Chapelle and L.~Li, ``An empirical evaluation of {T}hompson sampling,'' in
  \emph{Proc. of the 24th Int. Conf. on Neural Information Processing Systems},
  Granada, Spain, 2011, pp. 2249--2257.

\bibitem{MiWEBA}
\BIBentryALTinterwordspacing
A.~Maltsev, A.~Pudeyev, and et~al., ``{WP}5: Propagation, antenna, and
  multi-antenna techniques: D5.1 - channel modeling and characterization,'' EU
  and Japanese Government, Tech. Rep., Jun. 2014. [Online]. Available:
  \url{https://www.miweba.eu/wp-content/uploads/2014/07/MiWEBA_D5.1_v1.01.pdf}
\BIBentrySTDinterwordspacing

\bibitem{YiZhang_DeepRL_mmWave_MANET_Codes}
\BIBentryALTinterwordspacing
Y.~Zhang and R.~W. Heath. (2021) Source code for paper titled reinforcement
  learning-based joint user scheduling and link configuration in
  millimeter-wave networks. [Online]. Available:
  \url{https://github.com/yzhang417/DeepRL-mmWave-MANET/}
\BIBentrySTDinterwordspacing

\end{thebibliography}

\begin{IEEEbiography}
	[{\includegraphics[width=1in,height=1.25in,clip,keepaspectratio]{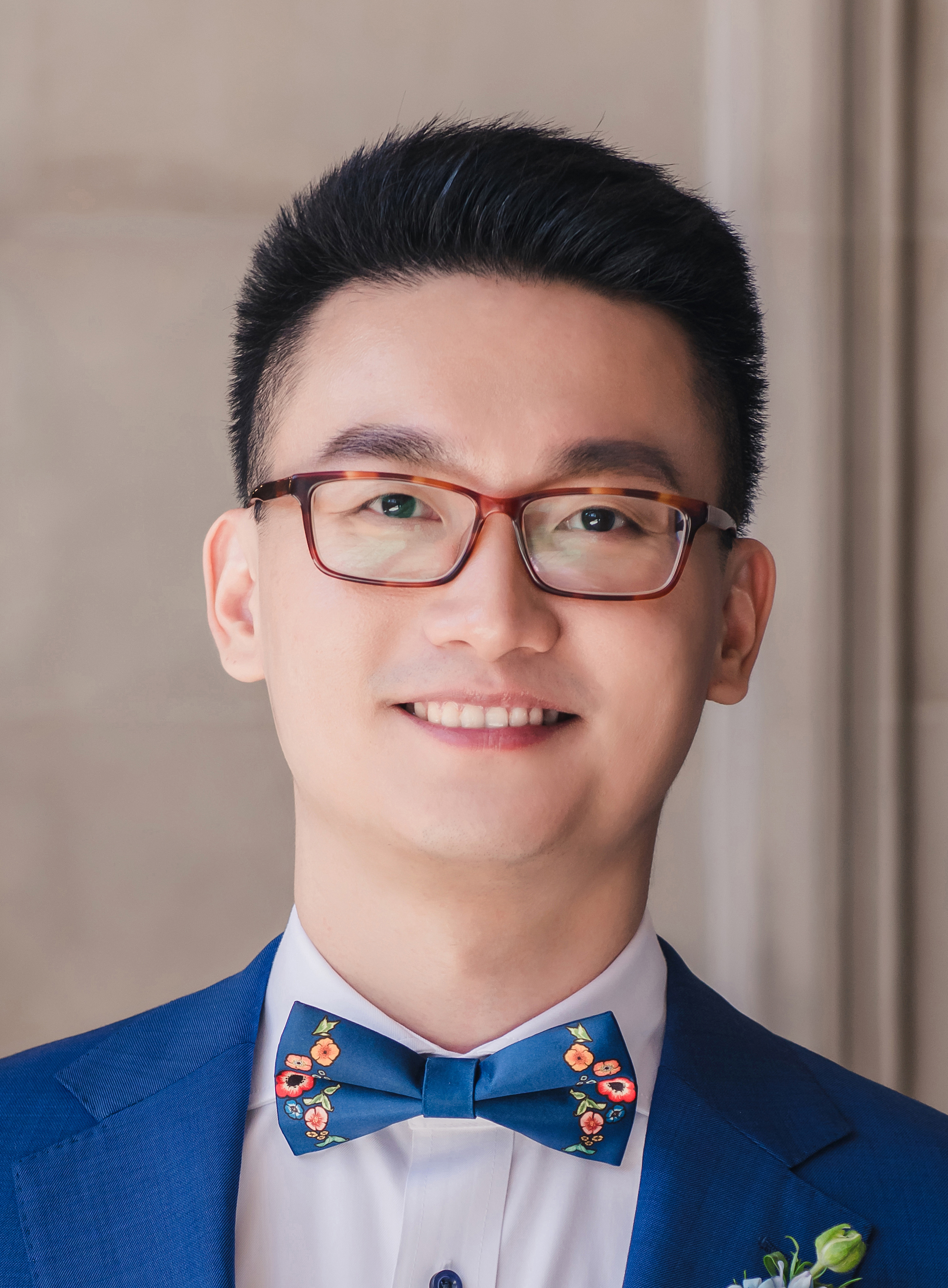}}]
	{Yi Zhang} (S'15-M'21) Yi Zhang received his Ph.D. degree in Electrical and Computer Engineering from The University of Texas at Austin in 2021. His dissertation supervisors are Prof. Robert Heath and Prof. Sanjay Shakkottai. Before that, he received B.S./M.S. degrees from Xi'an Jiaotong University in 2014 and 2017, respectively. He also received an Engineer's degree from Ecole Centrale de Nantes, France, in 2017. His research interests include wireless PHY/MAC design, WLAN/Cellular, signal processing, deep learning, and reinforcement learning (DRL and bandit). He is currently a wireless system engineer at Apple Inc in Sunnyvale, CA.
\end{IEEEbiography}

\begin{IEEEbiography}
	[{\includegraphics[width=1in,height=1.25in,clip,keepaspectratio]{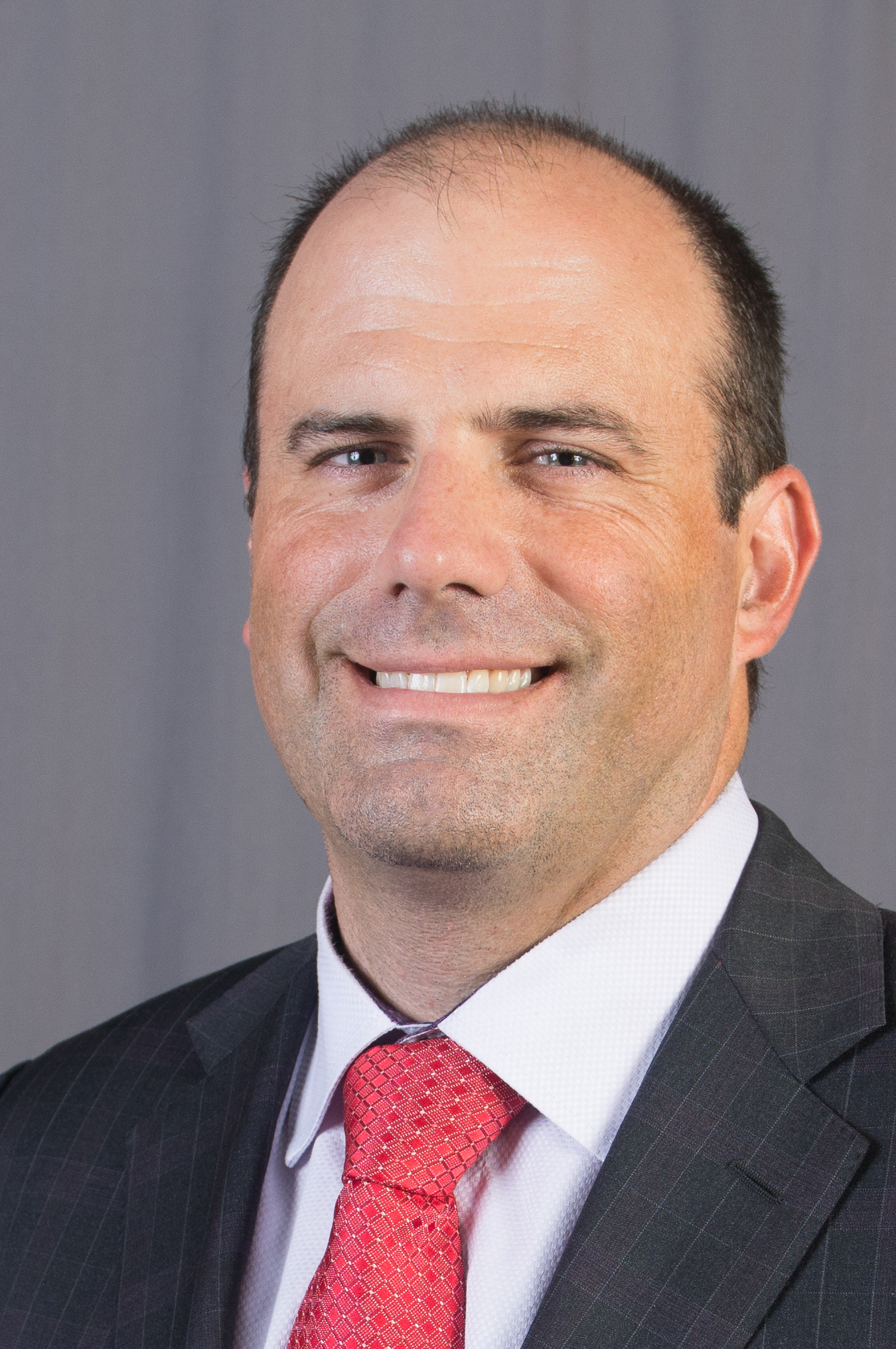}}]
	{Robert W. Heath Jr.} (S'96-M'01-SM'06-F'11)  received the B.S. and M.S. degrees from the University of Virginia, Charlottesville, VA, in 1996 and 1997 respectively, and the Ph.D. from Stanford University, Stanford, CA, in 2002, all in electrical engineering. From 1998 to 2001, he was a Senior Member of the Technical Staff then a Senior Consultant at Iospan Wireless Inc, San Jose, CA where he worked on the design and implementation of the physical and link layers of the first commercial MIMO-OFDM communication system. He is a Distinguished Professor at North Carolina State University. From 2002-2020 he was with The University of Texas at Austin, most recently as Cockrell Family Regents Chair in Engineering and Director of UT SAVES. He is also President and CEO of MIMO Wireless Inc. He authored ``Introduction to Wireless Digital Communication'' (Prentice Hall, 2017) and ``Digital Wireless Communication: Physical Layer Exploration Lab Using the NI USRP'' (National Technology and Science Press, 2012), and co-authored ``Millimeter Wave Wireless Communications'' (Prentice Hall, 2014) and ``Foundations of MIMO Communication'' (Cambridge University Press, 2018). He is currently Editor-in-Chief of IEEE Signal Processing Magazine.
	
	Dr. Heath has been a co-author of a number award winning conference and journal papers including recently the 2016 IEEE Communications Society Fred W. Ellersick Prize, the 2016 IEEE Communications and  Information Theory Societies Joint Paper Award, the 2017 Marconi Prize Paper Award, and the 2019 IEEE Communications Society Stephen O. Rice Prize. He received the 2017 EURASIP Technical Achievement award and the 2019 IEEE Kiyo Tomiyasu Award. He was a distinguished lecturer and member of the Board of Governors in the IEEE Signal Processing Society. In 2017, he was selected as a Fellow of the National Academy of Inventors. He is also a licensed Amateur Radio Operator, a Private Pilot, a registered Professional Engineer in Texas. 
\end{IEEEbiography}

\end{document}